\documentclass[final,1p,times,authoryear]{elsarticle}

\usepackage{amssymb}
\usepackage{amsmath}

\usepackage{algorithm}
\usepackage{savesym}
\savesymbol{AND}
\usepackage{algorithmic}

\usepackage{graphicx}
\usepackage{amsthm}
\usepackage[dvipsnames]{xcolor}
\usepackage{amsmath,bm}

\usepackage{hyperref}
\hypersetup{
    colorlinks = true,
    linkcolor = blue,
    filecolor = magenta,      
    urlcolor = cyan,
    citecolor = Blue,
    }
\newtheorem{theorem}{Theorem}
\newtheorem{lemma}{Lemma}
\newtheorem{example}{Example}%
\newtheorem{remark}{Remark}%

\newtheorem{definition}{Definition}%
\usepackage[caption=false,font=normalsize,labelfont=sf,textfont=sf]{subfig}

\usepackage{tikz}
\usetikzlibrary{arrows.meta,bending,positioning,shapes.geometric}

\usetikzlibrary{decorations,decorations.markings} 
\pgfkeys{/tikz/.cd,
    contour distance/.store in=\ContourDistance,
    contour distance=-10pt, 
    contour step/.store in=\ContourStep,
    contour step=1pt,
}

\pgfdeclaredecoration{closed contour}{initial}
{%
\state{initial}[width=\ContourStep,next state=cont] {
    \pgfmoveto{\pgfpoint{\ContourStep}{\ContourDistance}}
    \pgfcoordinate{first}{\pgfpoint{\ContourStep}{\ContourDistance}}
    \pgfpathlineto{\pgfpoint{0.3\pgflinewidth}{\ContourDistance}}
    \pgfcoordinate{lastup}{\pgfpoint{1pt}{\ContourDistance}}
    
  }
  \state{cont}[width=\ContourStep]{
     \pgfmoveto{\pgfpointanchor{lastup}{center}}
     \pgfpathlineto{\pgfpoint{\ContourStep}{\ContourDistance}}
     \pgfcoordinate{lastup}{\pgfpoint{\ContourStep}{\ContourDistance}}
  }
  \state{final}[width=\ContourStep]
  { 
    \pgfmoveto{\pgfpointanchor{lastup}{center}}
    \pgfpathlineto{\pgfpointanchor{first}{center}}
  }
}

\usepackage{pgfplots}
\pgfplotsset{compat=1.18}

\newcommand{\m}[1]{\mathbf{#1}}
\newcommand{\mc}[1]{\mathcal{#1}}
\newcommand{\mb}[1]{\mathbb{#1}}

\journal{\cdots}

\begin{document}

\begin{frontmatter}



\title{Warshall algorithm for matrix-weighted graphs} 


\author{Minh Hoang Trinh\corref{cor1}} 

\affiliation{organization={Tayan Polytechnic},
            addressline={Tan Hoi, O Dien}, 
            city={Hanoi},
            postcode={10000},
            country={Vietnam}
            }

\author{Hyo-Sung Ahn} 
%
\affiliation{organization={School of Mechanical Engineering, Gwangju Institute of Science and Technology},
           addressline={123 Cheomdangwagi-ro, Buk-gu}, 
           city={Gwangju},
           postcode={500-712}, 
           country={Republic of Korea}
           }
\cortext[cor1]{Corresponding author. E-mail: \texttt{minhtrinh@ieee.org}}      
\begin{abstract}
This paper proposes Warshall-type algorithms for determining connectedness and clustering in an undirected matrix-weighted graphs. While a path between two vertices guarantees their connectedness in a scalar-weighted graph, the existence of one or more paths between them does not necessarily guarantee that they belong to the same cluster in a matrix-weighted graph. First, a sufficient condition for pairwise connectedness is established via aggregating path kernels between them. Second, we introduce three block matrix logic operators that enables the connectedness condition to be compactly represented and manipulated with positive semidefinite matrices. The proposed Warshall algorithm simultaneously determines connectivity between every pair of vertices in the graph and provides an approximated graph partition. Third, a distributed version of the Warshall algorithm is developed. Proofs of correctness, together with computational complexity analysis and numerical examples, are provided to establish the validity of the proposed algorithms.
\end{abstract}




\end{frontmatter}



\section{Introduction}
\label{sec:intro}
A matrix-weighted graph is a multidimensional generalization of a classical graph, where symmetric positive semidefinite matrices are associated with edges rather than positive scalars. Notably, matrix-weighted graphs are capable of representing intra-and cross-layer interactions in multidimensional networks \citep{Kivela2014,Trinh2024networked,Vu2025consensus}, clustering phenomenon arising under linear diffusion dynamics \citep{Trinh2018matrix}, as well as providing a unified representation for the stiffness and bearing Laplacian matrices \citep{Zelazo2015decentralized,Trinh2016cdc,Zhao2016localizability}. Several properties of matrix-weighted graphs with positive definite matrix weights have been introduced in \citep{Barooah2005distributed,Barooah2008estimation}. Physical interpretations of matrix-weighted graphs in circuit theory \citep{Atik2019resistance,Mahato2023squared,De2020H2}, and applications in consensus and clustering dynamics \citep{Tran2021discrete,Miao2021second}, multidimensional opinion dynamics \citep{Trinh2017ASCC,Ahn2020opinion,Pan2018bipartite}, privacy enhancement consensus protocols \citep{Pan2025privacy}, formation control and network localization \citep{Barooah2008estimation,Zhao2016localizability} were also considered in the literature. As a further abstraction, \citep{Hansen2021expansion} studied various abstract properties of the matrix-weighted Laplacian via cellular sheaf theory.

Connectivity is a fundamental property of networks and has been extensively studied in the context of scalar-weighted graphs. A connectivity algorithm partitions an undirected network into clusters (or classes) via vertex transitive closure property, i.e., the existence of a path between two vertices ensures connectedness between them \citep{Roy1959,Warshall1962}. Clustering information can be used to predict the network's behaviors and characteristics, for example, the asymptotic behavior of the network's state under a diffusion dynamics, or under a consensus algorithm-a mainstream research of control systems society over the recent decades. 

Investigating connectedness and clusters in a matrix-weighted graph, however, is not a trivial task. As the edges are weighted by positive semidefinite matrices, the existence of several paths joining two vertices do not guarantee they belong to a same cluster. Precise algorithms for determining connectivity between vertices of a matrix-weighted graph were proposed in \citep{Trinh2026corrigendum,Trinh2025} based on the definiteness of the effective resistance matrix \citep{Trinh2026corrigendum} or the grounded Laplacian \citep{Trinh2025}. These algorithms were developed analogously to the corresponding concepts of electrical networks \citep{Doyle1984random,Spielman2011,Nachmias2020}. However, exact approaches are computational demanding and does not exploit any information about the graph topology. Several approximated approaches in analysing connectedness and clustering of matrix weighted graphs were also proposed in the literature. For examples, a greedy algorithm was introduced in \citep{Trinh2018matrix} which partitions a matrix-weighted graph into subgraphs (clusters) according to its positive trees and subsequently combines these subgraphs based on algebraic conditions along the paths connecting them. The authors in \citep{Kwon2020TCNS} provided several conditions for clustering and consensus in each dimension of the matrix-weighted graph. Also, the authors in \citep{Pan2020CSL} estimated the rank of the controllable subspace of the matrix-weighted Laplacian with regard to an input matrix and provided specific controllability condition for positive paths or cycle graphs.

This paper presents several heuristic algorithms to determine whether a matrix-weighted graph is connected and, if not, to provide a reasonable partition of its vertices into clusters. We first prove a sufficient condition for two vertices to be \emph{connected}, i.e., the kernel of the matrix-weighted Laplacian contains only vectors whose subvectors associated with these vertices are always equal. We consider each positive semidefinite matrix weight as a generalized conductance, and the whole graph as a  conductance network. The connectivity assessment between two vertices can be divided into smaller kernel determination problems for a set of paths joining them, and the paths' kernels are then combined to derive a conclusion for the assessment. The divide-and-conquer strategy improves a brute-force search algorithm in \citep{Trinh2018matrix}, which was proposed to determine all clusters from an initial partition of a matrix-weighted graph into positive trees. Based on the connectivity condition, we introduce Warshall algorithm for undirected matrix-weighted graphs. The proposed algorithm  can be considered as a parallel path kernel computation with shared memory, which simultaneously determines the transitive closures of $n$ vertices. In designing the Warshall algorithm, the notions of parallel and series matrix additions (``$\vee$'' and ``$\wedge$'' logic operators) are introduced for block matrices,\footnote{Parallel and series matrix additions were defined in \citep{Anderson1969Matrices} for positive semidefinite matrices.} which enables a compact description of kernel computations in parallel. To further reduce the computational complexity, a novel decision operator is also defined and employed in each immediate step of the Warshall algorithm. Third, a distributed version of the newly developed Warshall algorithm is also developed. The distributed algorithm assumes that the $n$ distributed computing agents have access only to local graph information, thereby eliminating the need for shared memory while enhancing information privacy. By utilizing $n$ parallel iterations of the Warshall algorithm, each agent can determine the vertices in the same cluster heuristically, and with a lowered computational complexity. The idea of splitting the problem into $n$ sub-problem and combining the end result in the distributed Warshall algorithm is closely related to the SO$(d)$ orientation estimation algorithm presented in \citep{Lee2018distributed}. It is also worth remarking that distributed implementation of Warshall algorithm for scalar-weighted graph was developed in \citep{vanDeSnepscheut1986}. 
Although the Warshall algorithm was introduced in \citep{Trinh2025}, there has been no formal analysis. This work provides proofs of correctness and computational complexity analyses for both the Warshall algorithm and its distributed version.

The remainder of this paper are structured as follows.  Section~\ref{sec:prel} contains a preliminaries on matrix-weighted graphs and parallel and series additions of symmetric positive semidefinite matrices. Pairwise connectivity conditions, Warshall algorithm, and distributed Warshall algorithm are studied in Section~\ref{sec:alg}. Section~\ref{sec:examples} contains several examples illustrating the Warshall algorithm, and Section~\ref{sec:conclusion} concludes the paper.

\section{Preliminaries}
\label{sec:prel}

\subsection{Matrix-weighted graphs}
Consider an undirected matrix-weighted graph $G=(V,E,W)$, where $V=\{v_1,\ldots,v_n\}$ is the set of $|V|=n$ vertices, $E\subset V\times V$ is the set of $|E|=m$ edges, and $W=\{\m{A}_{ij}\in \mb{R}^{d\times d}|(v_j,v_i)\in E\}$ is the set of matrix weights. We assume $G$ does not have self-loops and multi-edge. The matrix weights are symmetric positive semidefinite, and as the graph is undirected,  the matrix weights satisfy $\m{A}_{ij}=\m{A}_{ji}=\m{A}_{ij}^\top$ for all $(v_j,v_i)\equiv (v_i,v_j)\in E$. If an edge $(v_j,v_i)$ has $\m{A}_{ij}$ positive definite, then it is called a positive definite edge.

If $(v_j,v_i)\in E$, then two vertices $v_i$ and $v_j$ are adjacent to each other. The neighbor set of a vertex $v_i$ can be defined as $\mc{N}_i=\{v_j\in V| (v_j,v_i) \in E\}$. A \emph{complete graph} has $\mc{N}_i = V \setminus \{v_i\}, \forall i \in V$. A \emph{path} in $G$ is defined as a sequences of edges joining adjacent vertices in $G$. For example, the path $\mc{P}=v_{i1}v_{i2}\ldots v_{il}$ joins the start vertex $v_{i1}$ to the end vertex $v_{il}$ traverses through edges $(v_{ik},v_{i,k+1})\in E$, $k=1,\ldots,l-1$. The length of $\mc{P}$ equals the number of edges in $\mc{P}$, which is $|\mc{P}|=l-1$ in this example. The path $\mc{P}$ is a simple path if and only if it contains no repeated vertices. A \emph{cycle} is a path with the same starting and ending vertices.

Corresponding to the matrix-weighted graph $G$, we can define the \emph{topological graph} $(V,E)$. The matrix-weighted graph $G$ is \emph{topologically connected} if and only if for any pairs of vertices in $V$, there exists a path in $(V,E)$ joining them. The concept of \emph{connectedness} in a matrix-weighted graph, however, requires both topological connectedness and algebraic conditions related to the matrix-valued weights. We label the edges in $E$ as $e_1,\ldots,e_m$ and for each edge $e_k = (v_i,v_j)\in E$, a vertex is chosen as the start vertex and the other is the end vertex of the edge. Correspondingly, the incident matrix $\m{H}=[h_{ki}] \in \mb{R}^{m \times n}$ of the graph can be defined with
\begin{align*}
h_{ki} = \left\lbrace \begin{array}{rl}
1,  & \text{if vertex } v_i \text{ is the starting vertex of } e_k, \\
-1, & \text{if vertex } v_i \text{ is the ending vertex of } e_k, \\
0,  & \text{otherwise.}
\end{array}
\right.
\end{align*}

The matrix-weighted adjacency matrix $\m{A} = [\m{A}]_{ij} \in \mb{R}^{dn\times dn}$ of $G$ is a block matrix, with the $i,j$-th sub-blocks $[\m{A}]_{ij} = \m{A}_{ij}$. Let $\m{D}_i = \sum_{j \in \mc{N}_i}\m{A}_{ij}$ be the degree of a vertex $v_i \in V$, the degree matrix of $G$ is a block diagonal matrix with matrices $\m{D}_i$ in the main diagonal, i.e., $\m{D}={\rm blkdg}(\m{D}_1,\ldots,\m{D}_n)$. The matrix-weighted Laplacian of $G$ is defined as $\m{L} = \m{D} - \m{A}$. The matrix-weighted Laplacian can also be expressed as $\m{L}=({\m{H}}^\top \otimes \m{I}_d)\m{W} ({\m{H}} \otimes \m{I}_d)$, where ``$\otimes$'' denotes the Kronecker product, $\m{I}_d$ denotes the identity matrices of size $d$, and $\m{W}={\rm blkdg}(\ldots,\m{A}_{ij},\ldots) \in \mb{R}^{dm \times dm}$ is the block diagonal matrix with matrices $\m{A}_{ij}$ in the main diagonal and being placed in the same order as we label the edges in the incidence matrix $\m{H}$. 

We can now define the concepts of connectedness and clustering in a matrix-weighted graph based on its corresponding matrix-weighted Laplacian \citep{Trinh2018matrix}.

\begin{definition}[Connected and clustering] \label{def:connectedness_clustering} Let $G$ be an undirected matrix-weighted graph with the matrix-weighted Laplacian $\m{L}$. Then, $G$ is connected if and only if $\text{rank}(\m{L})=dn-d$. Otherwise, $G$ is clustering.
\end{definition}

Clearly, topologically connectedness is necessary but not sufficient for a matrix-weighted graph to be connected. Thus, in this work, all considered matrix-weighted graphs are assumed to be topologically connected. 

Since connectedness implies the existence of a cluster containing every pair of vertices, we have the following definition.

\begin{definition}[Pairwise vertex connectedness]\label{def:same_cluster}
Let $G$ be an undirected matrix-weighted graph with the matrix-weighted Laplacian $\m{L}$. Two vertices $v_i \ne v_j$ belong to a same cluster, or connected, if and only if for any vector $\m{x}=[\m{x}_1^\top,\ldots,\m{x}_n^\top]^\top\in \mb{R}^{dn}$ in ${\rm ker}(\m{L})$, there holds $\m{x}_i = \m{x}_j$.
\end{definition}

Before devising new algorithms for testing connectedness and clustering in matrix-weighted graphs in Section~\ref{sec:alg}, we introduce matrix logic operators in subsection~\ref{subsec:matrix_operators}. These matrix operators are elementary for stating these algorithms.

\subsection{The ``$\vee$'' and ``$\wedge$'' operators}
\label{subsec:matrix_operators}
Let the matrices $\m{A}_k \in \mb{R}^{d\times d}$, $k=1,2,\ldots,N$, $d\geq 2$ be symmetric positive semidefinite, i.e., $\m{A}_k \succeq 0$. 

A matrix-weighted graph can be physically interpreted as a generalized conductance circuit. Thus, it is natural to deduce composition rules of the matrix conductances \citep{Anderson1969Matrices}. The \emph{series addition}, or the ``$\vee$'' operator, of two symmetric positive semidefinite matrices $\m{A}_1$ and $\m{A}_2$ is defined as $\m{A}_1\vee\m{A}_2 = \m{A}_1 + \m{A}_2$. Similarly, $\bigvee_{i=1}^N \m{A}_k = \left( \left(\m{A}_1 \vee \m{A}_2 \right)\vee \ldots \right) \vee \m{A}_N$ implies that the ``$\vee$'' operators are sequentially applied.

The \emph{parallel addition}, or the ``$\wedge$'' operator, of two matrices $\m{A}_1$ and $\m{A}_2$ is defined as 
\begin{align} \label{eq:parallel_sum}
\m{A}_1 \wedge \m{A}_2 = \m{A}_1(\m{A}_1 + \m{A}_2)^{\dagger}\m{A}_2,
\end{align}
where $(\m{A}_1 + \m{A}_2)^{\dagger}$ denotes the Moore-Penrose pseudo-inverse of $\m{A}_1 + \m{A}_2$. In a similar manner, we write $\bigwedge_{k=1}^N \m{A}_k =\left( \left(\m{A}_1 \wedge \m{A}_2 \right) \wedge \ldots \right) \wedge \m{A}_N$ to imply that the ``$\wedge$'' operators are sequentially applied.

With ${\rm ker}$ denotes the kernel space, we have the following lemmas on the ``$\vee$'' and ``$\wedge$'' operators whose proofs can be found in \citep{Anderson1969Matrices}. Since the claims of Lemma \ref{lem:parallel_sum} is not straightforward, a new and simpler proof is provided in \ref{app:parallel_sum}.

\begin{lemma}[The ``$\vee$'' operator] \label{lem:series_sum} 
The ``$\vee$'' operator of $d\times d$ symmetric positive semidefinite matrices satisfies the following properties:
\begin{itemize}
\item[i.] Commutativity: $\m{A}_1\vee \m{A}_2 = \m{A}_2\vee\m{A}_1$,
\item[ii.] Symmetric positive semidefinite: $\m{A}_1\vee \m{A}_2 = (\m{A}_1\vee \m{A}_2)^\top \succeq 0$,
\item[iii.] Kernel space: ${\rm ker}(\m{A}_1\vee \m{A}_2) = {\rm ker}(\m{A}_1) \cap {\rm ker}(\m{A}_2)$,
\item[iv.] Associativity: $(\m{A}_1\vee \m{A}_2)\vee \m{A}_3 = \m{A}_1 \vee (\m{A}_2 \vee \m{A}_3) = \m{A}_1 \vee \m{A}_2 \vee \m{A}_3$.
\end{itemize}
\end{lemma}

\begin{lemma}[The ``$\wedge$'' operator] \label{lem:parallel_sum}
The ``$\wedge$'' operator of $d\times d$ symmetric positive semidefinite matrices satisfies the following properties:
\begin{itemize}
\item[i.] Commutativity: $\m{A}_1\wedge\m{A}_2 = \m{A}_2\wedge\m{A}_1$,
\item[ii.] Symmetric positive semidefinite: $\m{A}_1 \wedge \m{A}_2 = (\m{A}_1 \wedge \m{A}_2)^\top \succeq 0$,
\item[iii.] Kernel space: ${\rm ker}(\m{A}_1\wedge\m{A}_2) = {\rm ker}(\m{A}_1) + {\rm ker}(\m{A}_2)=\{\m{u}+\m{v}\mid \m{u}\in {\rm ker}(\m{A}_1),\m{v}\in{\rm ker}(\m{A}_2)\}$, 
\item[iv.] Associativity: ${\rm ker}((\m{A}_1\wedge\m{A}_2)\wedge\m{A}_3) = {\rm ker}(\m{A}_1\wedge(\m{A}_2\wedge\m{A}_3)) = {\rm ker}(\m{A}_1 \wedge \m{A}_2 \wedge \m{A}_3)$,
\item[v.] Distributivity: ${\rm ker}((\m{A}_1\vee\m{A}_2)\wedge\m{A}_3) \subseteq {\rm ker}\left((\m{A}_1\wedge \m{A}_3) \vee(\m{A}_2\wedge\m{A}_3) \right)$.
\end{itemize}
\end{lemma}

In particular, we have ${\rm ker}(\m{A} \wedge \m{A})={\rm ker}(\m{A})$, ${\rm ker}(\m{A} \wedge \bm{\Theta}_d) = \mb{R}^d$, ${\rm ker}(\m{A} \vee \m{A})={\rm ker}(\m{A})={\rm ker}(\m{A} \vee \bm{\Theta}_d)$. 

\section{Algorithms}
\label{sec:alg}
As aforementioned in the previous section, the ``$\wedge$'' and ``$\vee$'' operators offer quantitative tools to evaluate the connectivity. In this section, we first devise sufficient conditions for pairwise connectedness in an undirected matrix-weighted graph. Second, the proposed condition will be exploited to develop Warshall algorithm for determining connectedness and clustering in a given matrix-weighted graph. Finally, we propose a distributed version of the Warshall algorithm, which shares the computational load among a network of computing agents.

\subsection{Sufficient conditions for pairwise connectedness}
\label{subsec:BFS}
For each pair of vertices $v_i\ne v_j$, $v_i,v_j\in V$, let
\begin{align}
\mc{S}_{ij}^{\infty} &= \{\mc{P}_k=v_1^k\ldots v_{|\mc{P}_k|+1}^k|~v_1^k = v_i, v_{|\mc{P}_k|+1}^k=v_j\}, \\
\mc{S}_{ij} &= \{\mc{P}_k\in \mc{S}_{ij}^{\infty}|~\mc{P}_k \text{ is simple}\},
\end{align}
be respectively the set of all paths from $v_i$ to $v_j$ and the set of all simple paths from $v_i$ to $v_j$. 

\begin{figure*}[t!]
\centering
{\begin{tikzpicture}[
roundnode/.style={circle, draw=black, thick, minimum size=3.5mm,inner sep= 0.25mm},
squarednode/.style={rectangle, draw=black, thick, minimum size=3.5mm,inner sep= 0.25mm},
]
    \node[roundnode] (u1) at (0,0) { }; %
    \node[roundnode] (u2) at (1,1.15) { };%
    \node[roundnode] (u3) at (2.5,1.7) { };%
    \node[roundnode] (u4) at (4.5,1) { };%
    \node[roundnode] (u5) at (5.5,0) { };%
    \node[roundnode] (u6) at (1.5,0) { };%
    \node[roundnode] (u7) at (4,0) { };%

    \node[roundnode] (v2) at (1,-1.15) { };%
    \node[roundnode] (v3) at (2.5,-1.7) { };%
    \node[roundnode] (v4) at (4.5,-1) { };%
    
    \draw[-, very thick] (2,0)--(u6)--(u1)--(u2);
    \draw[-, very thick] (u2)--(u3);
    \draw[-, very thick] (u3)--(3,1.6);
    \draw[-, very thick] (u4)--(u5);
    \draw[-, very thick] (u4)--(4,1.2);
    \draw[-, very thick] (u5)--(u7)--(3.5,0);
    
    \draw[-, very thick] (u1)--(v2)--(v3);
    \draw[-, very thick] (v3)--(3,-1.6);
    \draw[-, very thick] (v4)--(u5);
    \draw[-, very thick] (v4)--(4,-1.2);
    
    \node (l1) at (3.5,1.5) {\Large $\ddots$};
    \node (r1) at (3.5,-1.3) {\Large \reflectbox{$\ddots$}};
    \node (l2) at (2.8,0) {\Large $\ldots$};
    \node (l3) at (2.8,.8) {\Large $\vdots$};
    \node (r3) at (2.8,-.7) {\Large $\vdots$};
    \node (t1) at (-0.5,-0.5) {$v_{i_1}^k \equiv v_i$};
    \node (t2) at (0.55,1.45) {$v_{i_2}^k$};
    \node (p) at (4,1.8) {\Large $\mc{P}_k$};
    \node (t3) at (2.5,2.2) {$v_{i_3}^k$};
    \node (t4) at (5.1,1.25) {$v_{i_{|\mc{P}_k|}}^k$};
    \node (t5) at (6,-0.5) {$v_j \equiv v_{i_{|\mc{P}_k|+1}}^k$};
    
    \node[roundnode] (w1) at (-0.5,1) { }; %
    \node[roundnode] (w2) at (-1.5,1.25) { }; %
    \node[roundnode] (w3) at (-2.5,0.5) { }; %
    \node[roundnode] (w4) at (-1.25,0) { }; %
    \draw[-, very thick] (u1)--(w1)--(w2);
    \draw[-, very thick] (u1)--(w4)--(w3);
    \node (q1) at (-1.95,1) {\Large \reflectbox{$\ddots$}};
    \node (q2) at (-1.2,0.55) {\Large $\mc{C}^k_1$};
    
    \node[roundnode] (x1) at (1.5,2.5) { }; %
    \node[roundnode] (x2) at (2.25,3) { }; %
    \node[roundnode] (x3) at (3.3,2.3) { }; %
    \draw[-, very thick] (u3)--(x1)--(x2);
	\draw[-, very thick] (u3)--(x3);
	\node (q3) at (2.75,2.8) {\Large {$\ddots$}};
	\node (q4) at (3.1,2.9) {\Large $\mc{C}^k_2$};
	
	\node[roundnode] (y1) at (6,1) { }; %
    \node[roundnode] (y2) at (7,1.25) { }; %
    \node[roundnode] (y3) at (8,0.5) { }; %
    \node[roundnode] (y4) at (6.75,0) { }; %
    \draw[-, very thick] (u5)--(y1)--(y2);
    \draw[-, very thick] (u5)--(y4)--(y3);
    \node (q5) at (7.6,1) {\Large {$\ddots$}};
    \node (q6) at (6.8,0.6) {\Large $\mc{C}^k_3$};

   \draw plot[smooth cycle, thick] coordinates {(-0.3,0.1) (0.8,1.3) (2.45,1.95) (4.5,1.3) (5.75,0.1) (5.4,-0.2) (4.3,0.8) (2.7,1.35) (1.15,.85) (0.1,-0.25)};
\end{tikzpicture}}
\caption{The path $\mc{P}_k=v_{i_1}^kv_{i_2}^k\ldots v_{i_{|\mc{P}_k|+1}}^k$ joins two vertices $v_i$ and $v_j$.}
\label{fig:paths}
\end{figure*}

For each path $\mc{P}_k$ (see Fig.~\ref{fig:paths} for an illustration), we define 
\begin{align} \label{eq:ker_Pk}
{\rm ker}(\mc{P}_k) \triangleq \sum_{j=1}^{{|\mc{P}_k|}} {\rm ker}\left(\m{A}_{v^k_jv^k_{j+1}}\right) = {\rm ker}\left(\bigwedge_{j=1}^{|\mc{P}_k|}\m{A}_{v_j^kv_{j+1}^k}\right).
\end{align}

By sorting the elements in the sets $\mc{S}_{ij}$ and $\mc{S}_{ij}^\infty$ by their lengths, we can decompose them into subsets $\mc{S}_{ij}^t$, $t=1,2,\ldots,$ containing all paths of length $t$ from $v_i$ to $v_j$. Since $\mc{S}_{ij}$ only admits simple paths, the lengths of every element in $\mc{S}_{ij}$ is at most $n-1$.

Let
\begin{align}
{\rm ker}(\mc{S}_{ij}^t) &= \bigcap_{\mc{P}_k\in\mc{S}_{ij}^t} {\rm ker}\left( \mc{P}_k\right)= {\rm ker}\left(\bigvee_{k=1}^{|\mc{S}_{ij}^t|} \bigwedge_{j=1}^{t}\m{A}_{v_j^kv_{j+1}^k}\right), \nonumber \\
{\rm ker}(\mc{S}_{ij}) &= \bigcap_{t=1}^{n-1}{\rm ker}(\mc{S}_{ij}^t), ~
{\rm ker}(\mc{S}_{ij}^{\infty}) = \bigcap_{t=1}^{\infty}{\rm ker}(\mc{S}_{ij}^t). \nonumber
\end{align}

Consider an arbitrary path $\tilde{\mc{P}} \in \mc{S}_{ij}^{\infty}$ joining $v_i$ and $v_j$ of length exceeding $n-1$. Then, $\tilde{\mc{P}}$ must visit some vertices in $V$ more than one time before ending at $v_j$. Thus, $\tilde{\mc{P}}$ must contain a simple path of shortest (topological) length $t$ $(1\le t\le n-1)$, after removing all redundant edges in $E(\tilde{\mc{P}})\setminus E({\mc{P}})$.\footnote{For example, in Fig.~\ref{fig:paths}, we may consider $\tilde{\mc{P}}_k$ consisting the cycle $\mc{C}^k_1$ initiating from $v_i$ joining with the path $v_iv_{i_2}^kv_{i_3}^k$, the cycle $\mc{C}^k_2$, the path $v_{i_3}^k \ldots v_{i_{|\mc{P}_k|+1}}$, and the cycle $\mc{C}^k_3$. The vertices in each $\mc{C}^k_h,h=1,2,3$ may not be distinct and the repeated vertices may have appeared earlier in $\mc{P}_k$. After removing $\mc{E}(\mc{C}^k_1)$, $ \mc{E}(\mc{C}^k_2)$, $\mc{E}(\mc{C}^k_3)$ from $\tilde{\mc{P}}_k$, we obtain the simple path $\mc{P}_k$.}

Due to the commutative and associative properties of the wedge operator, we can write
\begin{align}
{\rm ker}(\tilde{\mc{P}}) = {\rm ker}({\mc{P}}) + {\rm ker}(E(\tilde{\mc{P}})\setminus E({\mc{P}})).
\end{align}
It follows that ${\rm ker}({\mc{P}}) \subseteq {\rm ker}(\tilde{\mc{P}})$, and ${\rm ker}(\mc{S}_{ij}^t) = {\rm ker}(\mc{S}_{ij}^t) \cap {\rm ker}(\tilde{\mc{P}})$. Therefore,
\begin{align} \label{eq:kernel_Sij_inf1}
{\rm ker}(\mc{S}_{ij}^\infty) = \bigcap_{t=1}^{\infty}{\rm ker}(\mc{S}_{ij}^t) = \bigcap_{t=1}^{n-1}{\rm ker}(\mc{S}_{ij}^t) = {\rm ker}(\mc{S}_{ij}).
\end{align}

The following lemma provides a sufficient condition for two vertices to belong to a cluster and a matrix weighted graph to be connected.

\begin{lemma}[Parallel path condition] \label{thm:connectedness} Consider an undirected matrix-weighted graph $G=(V,E,W)$ with the matrix-weighted Laplacian $\m{L}$. The following claims hold.
\begin{itemize}
\item[i.] Two vertices $v_i,v_j\in V$ belong to a same cluster if \begin{align} \label{eq:kernel_Sij}
{\rm dim}\left({\rm ker}(\mc{S}_{ij})\right) = 0.
\end{align}
\item[ii.] The graph is connected if equation~\eqref{eq:kernel_Sij} holds for every pair of distinct vertices $v_i, v_j\in V$.
\end{itemize}
\end{lemma}

\begin{proof}
i. Suppose that there are $|\mc{S}_{ij}| = q$ different paths labeled as $\mc{P}_1,\ldots,\mc{P}_q$ in the set $\mc{S}_{ij}$.

Consider a path $\mc{P}=v_{i_1}v_{i_2}\ldots v_{i_{|\mc{P}|}}v_{i_{|\mc{P}|+1}} \in \mc{S}_{ij}$, with $i_1 \equiv i$ and $i_{|\mc{P}|+1} \equiv j$. Consider an arbitrary vector $\m{x}=[\m{x}_1^\top,\ldots,\m{x}_n^\top]^\top\in \mb{R}^{dn}$, where each $\m{x}_i \in \mb{R}^d$ is associated with $v_i \in V$ such that $\m{L}\m{x} = \m{0}$.

Since ${\rm ker}(\m{L})={\rm ker}((\m{H}^\top\otimes \m{I}_d)\m{W}(\m{H}\otimes \m{I}_d)) = {\rm ker}(\m{W}(\m{H}\otimes \m{I}_d))$, based on \citep{Zelazo2010edge} , the restriction of the equation $\m{L}\m{x} = \m{0}_{dn}$ to the subgraph $\mc{P}$ is %
\begin{align}
\begin{bmatrix}
\m{A}_{i_1i_2} & 			    &        & \\
               &\m{A}_{i_2i_3}  &        & \\
               &                & \ddots & \\
               &                &        & \m{A}_{i_{|\mc{P}|}i_{|\mc{P}|+1}}
\end{bmatrix} (\m{H}_{\mc{P}}\otimes \m{I}_d) \m{x}= \m{0}_{d|\mc{P}|}, \label{eq:path-condition}
\end{align}
where $\m{H}_{\mc{P}}$ is the incident matrix corresponding to the path graph $\mc{P}$. 

Equivalently,
\begin{align} \label{eq:path-condition-subeq}
\m{A}_{i_ki_{k+1}}(\m{x}_{i_k} - \m{x}_{i_{k+1}}) = \m{0}_d,
\end{align}
for $k=1,\ldots,|\mc{P}|$. It is remarkable at this point that adding repeated subequations \eqref{eq:path-condition-subeq} does not alter solutions of \eqref{eq:path-condition}. Thus, equation \eqref{eq:path-condition} imposes no restriction that $\mc{P}$ must be a simple path. 

It follows that
\begin{subequations} \label{eq:x-path-P}
\begin{align}
\m{x}_{i_{2}} &= \m{x}_{i_{1}} + (\m{I}_d - \m{A}_{i_1 i_{2}}^{\dagger}\m{A}_{i_1 i_{2}})\m{y}_1, \\
&~\vdots \nonumber \\
\m{x}_{i_{|\mc{P}|+1}} &= \m{x}_{i_{|\mc{P}|}} + (\m{I}_d - \m{A}_{i_{|\mc{P}|} i_{|\mc{P}|+1}}^{\dagger}\m{A}_{i_{|\mc{P}|} i_{|\mc{P}|+1}})\m{y}_{|\mc{P}|},
\end{align}
\end{subequations}
for some arbitrary vectors $\m{y}_k \in \mb{R}^d$, $k=1,\ldots, |\mc{P}|$. Summing these equations \eqref{eq:x-path-P} side by side and eliminating common terms from both sides, we obtain
\begin{align*}
\m{x}_j - \m{x}_i &= \m{x}_{i_{|\mc{P}|+1}} - \m{x}_{i_{1}} = \sum_{r=1}^{|\mc{P}|} (\m{I}_d - \m{A}_{i_r i_{r+1}}^{\dagger}\m{A}_{i_r i_{r+1}})\m{y}_r.
\end{align*}
Since $(\m{I}_d - \m{A}_{i_r i_{r+1}}\m{A}_{i_r i_{r+1}}^{\dagger})\m{y}_r \in {\rm ker}(\m{A}_{i_r i_{r+1}}),\forall r = 1,\ldots, |\mc{P}|$, it follows that 
\begin{align} 
(\m{x}_{j} - \m{x}_{i}) \in \sum_{k=1}^{|\mc{P}|} {{\rm ker}(\m{A}_{i_{k}i_{i_{k+1}}})} = {\rm ker}(\mc{P}). \label{eq:each_kernel}
\end{align}
As there are $q$ paths $\mc{P}_k$ in $\mc{S}_{ij}$, by aggregating  $q$ equations as in \eqref{eq:each_kernel}, and using our assumption on ${\rm ker}(\mc{S}_{ij})$, we have
\begin{align}
(\m{x}_{j} - \m{x}_{i}) \in \bigcap_{k=1}^q {{\rm ker}(\mc{P}_k)} = {\bigcap_{k=1}^q {\rm ker}(\mc{P}_k)} = \{\m{0}_d\},
\end{align}
and this implies that $\m{x}_i = \m{x}_j$.

Therefore, if ${\rm dim}({\bigcap_{k=1}^q {\rm ker}(\mc{P}_k)}) = 0$, the kernel of  $\m{L}$ contains only vectors $\m{x} = [\m{x}_1^\top,\ldots,\m{x}_n^\top]^\top$ satisfying $\m{x}_i = \m{x}_j$, i.e., two vertices $v_i,v_j$ belong to a same cluster.

ii. This claim follows immediately from (i). 
\end{proof}


\begin{example}[A counter example] \label{eg:1}
Below, a counter example showing that condition \eqref{eq:kernel_Sij} is not necessary will be provided. Consider the matrix-weighted graph as depicted in Fig.~\ref{fig:graph-4} with
\begin{figure}[t!]
\centering
{\resizebox{6cm}{!}{\begin{tikzpicture}[
roundnode/.style={circle, draw=black, thick, minimum size=3.5mm,inner sep= 0.25mm},
squarednode/.style={rectangle, draw=black, thick, minimum size=3.5mm,inner sep= 0.25mm},
]
    \node[roundnode] (u1) at (0,0) {$v_1$}; %
    \node[roundnode] (u2) at (1,0) {$v_2$};%
    \node[roundnode] (u3) at (2,0) {$v_3$};%
    \node[roundnode] (u4) at (3,0) {$v_4$};%
    \node[roundnode] (u5) at (-2,0) {$v_5$};%
    \node[roundnode] (u6) at (-1,0) {$v_6$};%
    \draw[-, very thick] (u6)--(u1);
    \draw[-, very thick] (u1)--(u2)--(u3)--(u4);
    \draw[-, very thick] (u5)--(u6);
    \draw[very thick] (u4) edge [bend left=-30] (u5);
    \draw [very thick]
	(u3) edge [bend left=30] (u1);
\end{tikzpicture}}}
    \caption{The graph in Example~\ref{eg:1}}
    \label{fig:graph-4}
\end{figure}
\begin{align*}
\m{A}_{12} &=\m{A}_{23}= \begin{bmatrix}
1 & 1\\
1 & 1
\end{bmatrix},\m{A}_{45} = \m{A}_{56} = \begin{bmatrix}
1 & 0\\
0 & 0
\end{bmatrix},~
\m{A}_{13} = \begin{bmatrix}
1 & 2\\
2 & 4
\end{bmatrix}, \m{A}_{16} = \begin{bmatrix}
0 & 0\\
0 & 1
\end{bmatrix},\m{A}_{34}= \begin{bmatrix}
1 & 0\\
0 & 2
\end{bmatrix}.
\end{align*} 
Three paths between $v_1$ and $v_6$ and their corresponding kernels are $\mc{P}_1 = v_1v_6$, ${\rm ker}(\mc{P}_1)={\rm im}\left(\begin{bmatrix} 1 \\ 0 \end{bmatrix} \right)$, $\mc{P}_2 = v_1v_2v_3v_4v_5v_6$, ${\rm ker}(\mc{P}_2)=\mb{R}^{2}$, $\mc{P}_3 = v_1v_3v_4v_5v_6$, ${\rm ker}(\mc{P}_3)=\mb{R}^{2}$. 

Then, ${\rm ker}(\mc{S}_{16}) = \bigcap_{k=1}^3{\rm ker}(\mc{P}_k) = {\rm im}\left(\begin{bmatrix} 1 \\ 0 \end{bmatrix} \right)$. The connectivity condition~\eqref{eq:kernel_Sij} is not satisfied. 

The connectivity test \eqref{eq:kernel_Sij} satisfies for $v_1,v_3$ as \[{\rm ker}(v_1v_2v_4) \cap {\rm ker}(v_1v_3) = {\rm im}\left(\begin{bmatrix} 1 \\ -1 \end{bmatrix} \right) \cap {\rm im}\left(\begin{bmatrix} -2 \\ 1 \end{bmatrix} \right) = \{\m{0}_2\}.\] Identify $v_1$ and $v_3$ as a single vertex $v_{1,3}$, there are two paths from $v_{1,3}$ to $v_6$: $\mc{Q}_1=v_{1,3}v_6$ and $\mc{Q}_2 = v_{1,3}v_4v_5v_6$, with ${\rm ker}(\mc{Q}_1)={\rm im}\left(\begin{bmatrix} 1 \\ 0 \end{bmatrix} \right)$, ${\rm ker}(\mc{Q}_2)={\rm im}\left(\begin{bmatrix} 0 \\ 1 \end{bmatrix} \right)$, ${\rm ker}(\mc{Q}_1) \cap {\rm ker}(\mc{Q}_2) = \{\m{0}_2\}$. Thus, $v_1,v_3,v_6$ belong to the same cluster. This contradiction shows that condition~\eqref{eq:kernel_Sij} is not a necessary condition.
\end{example}

\begin{remark}[Relation to edge-connectivity] \label{rem:edge-connectivity}
It is remarked that the proof of Theorem~\ref{thm:connectedness} (i) involves all paths between $v_i$ and $v_j$, which hints a connection with the concept of $k$-connectedness in classical graph theory \citep{Diestel2025graph}. Menger's theorem states that a topological graph are $k$-connected if there exist $k$ vertex-disjoint paths between any two vertices. In matrix-weighted graphs, connectivity is jointly determined by the graph's topology (paths connecting vertices) and the values of the matrix weights. While there are many paths joining two vertices, if the matrix weights are not well chosen, they still do not belong to a same cluster. 

Meanwhile, it is not hard to show that if two vertices $v_i,v_j$ are joined by only a set of vertex-disjoint paths, the condition~\eqref{eq:kernel_Sij} is also necessary and sufficient condition for $v_i$ and $v_j$ to belong to a same cluster according to Definition~\ref{def:same_cluster}. 
\end{remark}

Since aggregating all paths joining $v_i,v_j$ in parallel requires searching for all paths joining $v_i,v_j$, using condition~\eqref{eq:kernel_Sij} would lead to an algorithm with exponential computational complexity. Below, we consider a new strategy for combining these paths based on intermediate vertices (see Figure~\ref{fig:paths}). 

\begin{figure}[t!]
\centering
{\begin{tikzpicture}[
roundnode/.style={circle, draw=black, thick, minimum size=3.5mm,inner sep= 0.25mm},
squarednode/.style={rectangle, draw=black, thick, minimum size=3.5mm,inner sep= 0.25mm},
]
    \node[roundnode] (u1) at (-0.5,0) { }; %
    \node[roundnode] (u2) at (2,0) { };%
    \node[roundnode] (u3) at (4,0) { };%
    
    \draw[-, very thick] (u1)--(0.2,0.7);
    \draw[-, very thick] (u1)--(0.2,0);
    \draw[-, very thick] (u1)--(0.2,-0.7);

    \draw[-, very thick] (u2)--(1.3,0.7);
    \draw[-, very thick] (u2)--(1.3,0);
    \draw[-, very thick] (u2)--(1.3,-0.7);        
    \draw[-, very thick] (u2)--(2.7,0.5);
    \draw[-, very thick] (u2)--(2.7,0);
    \draw[-, very thick] (u2)--(2.7,-0.5);    
    
    \draw[-, very thick] (u3)--(3.3,0.5);
    \draw[-, very thick] (u3)--(3.3,0);
    \draw[-, very thick] (u3)--(3.3,-0.5);    
    
    \node (l1a) at (.9,0.7) {\large $\cdots$};
	\node (l1b) at (.9,0.0) {\large $\cdots$};
	\node (l1c) at (.9,-0.7) {\large $\cdots$};

    \node (l2a) at (3,0.5) {$\cdots$};
	\node (l2b) at (3,0.0) {$\cdots$};
	\node (l2c) at (3,-0.5) {$\cdots$};

    \node (t1) at (-1,0) {$v_{i}$};
    \node (t2) at (2,-0.5) {$v_{k}$};
    \node (t3) at (4.5,0) {$v_j$};
    
    \node (p1) at (.75,1) { $\mc{P}_{ik,1}$};
    \node (p2) at (.75,-1) { $\mc{P}_{ik,r_1}$};
    
    \node (p3) at (3,.8) { $\mc{P}_{kj,1}$};
    \node (p4) at (3,-.8) { $\mc{P}_{kj,r_2}$};
    
\end{tikzpicture}}
\caption{The paths in $\{\mc{P}_{ik,r}\}_{r=1}^{r_1}$ join $v_i,v_k$ and the paths in $\{\mc{P}_{ik,r}\}_{r=1}^{r_1}$  join vertices $v_k,v_j$.}
\label{fig:path_common_point}
\end{figure}

\begin{lemma}[Intermediate vertex]\label{lem:path_reduction} Consider an undirected matrix-weighted graph $G$. Let $v_i \ne v_j$ are two distinct vertices and $v_k$ is an intermediate vertex. Let $\mc{P}_{ik,1},\ldots,\mc{P}_{ik,r_1}$ and $\mc{P}_{kj,1},\ldots,\mc{P}_{kj,r_2}$ be sets of paths joining $v_i$ to $v_k$ and $v_k$ to $v_j$, respectively. If 
\begin{align} \label{eq:reduction_lemma0}
{\rm dim}\left(\bigcap_{v_k\in V} \left(\bigcap_{l=1}^{r_2} \left( \bigcap_{r=1}^{r_1}{\rm ker}(\mc{P}_{ik,r}) + {\rm ker}(\mc{P}_{kj,l}) \right)\right)\right) = 0,
\end{align}
then $v_i,v_j$ belong to a same cluster.
\end{lemma}
\begin{proof}
Let $\m{x}=[\m{x}_1^\top,\ldots,\m{x}_n^\top]^\top \in \mb{R}^{dn}$ be an arbitrary vector in ${\rm ker}(\m{L})$. Applying Lemma~\ref{thm:connectedness} for $v_i$ and $v_k$ yields
\begin{align} \label{eq:reduction_lemma1}
\m{x}_i-\m{x}_k \in \bigcap_{r=1}^{r_1}{\rm ker}(\mc{P}_{ik,r}),
\end{align}
Similarly, for each $l=1,\ldots, r_2$,
\begin{align} \label{eq:reduction_lemma2}
\m{x}_k-\m{x}_j \in {\rm ker}(\mc{P}_{kj,r}).
\end{align}
It follows from \eqref{eq:reduction_lemma1} and \eqref{eq:reduction_lemma2} that 
\begin{align} \label{eq:reduction_lemma2a}
    \m{x}_i-\m{x}_j \in \bigcap_{r=1}^{r_1}{\rm ker}(\mc{P}_{ik,r})  + {\rm ker}(\mc{P}_{kj,l}).
\end{align}
Since \eqref{eq:reduction_lemma2a} holds for every $l=1,\ldots,r_2$, it turns out that
\begin{align} \label{eq:reduction_lemma2b}
    \m{x}_i-\m{x}_j \in \bigcap_{l=1}^{r_2} \left(\bigcap_{r=1}^{r_1}{\rm ker}(\mc{P}_{ik,r})  + {\rm ker}(\mc{P}_{kj,l})\right).
\end{align}
Furthermore, since \eqref{eq:reduction_lemma2b} holds for any $v_k \in V$, 
\begin{align}
\m{x}_i-\m{x}_j \in  \bigcap_{v_k\in V} \left(\bigcap_{l=1}^{r_2} \left( \bigcap_{r=1}^{r_1}{\rm ker}(\mc{P}_{ik,r}) + {\rm ker}(\mc{P}_{kj,l}) \right)\right),
\end{align}
and the lemma follows.
\end{proof}

Lemma~\ref{lem:path_reduction} is simple yet powerful, as it enables the kernels of multiple paths sharing a common intermediate vertex to be combined in a batch. Instead of computing kernels of $r_1r_2$ paths, we only need to determine $(r_1+r_2)+1$ path kernels.  The combination not only significantly reduces the computational complexity but also tightens the connectivity condition \eqref{eq:kernel_Sij}, owing to the fact that
 \begin{align}
 \bigcap_{r=1}^{r_1}{\rm ker}(\mc{P}_{ik,r}) &\subseteq {\rm ker}(\mc{P}_{ik,r}), \nonumber\\
 \bigcap_{r=1}^{r_1}{\rm ker}(\mc{P}_{ik,r}) + {\rm ker}(\mc{P}_{kj,l}) &\subseteq {\rm ker}(\mc{P}_{ik,r}) + {\rm ker}(\mc{P}_{kj},l) =  {\rm ker}(\mc{P}_{ik,r}\mc{P}_{kj,l}),~\forall r=1,\ldots,r_1, \nonumber \\
 \bigcap_{r=1}^{r_1}{\rm ker}(\mc{P}_{ik,r}) + {\rm ker}(\mc{P}_{kj,l}) &\subseteq \bigcap_{r=1}^{r_1}\left({\rm ker}(\mc{P}_{ik,r}\mc{P}_{kj},l)\right), \forall l=1,\ldots,r_2, \nonumber \\
  \bigcap_{l=1}^{r_2} \left(\bigcap_{r=1}^{r_1}{\rm ker}(\mc{P}_{ik,r}) + {\rm ker}(\mc{P}_{kj,l}) \right) &\subseteq \bigcap_{l=1}^{r_2} \bigcap_{r=1}^{r_1}{\rm ker}(\mc{P}_{ik,r}\mc{P}_{kj,l}). \label{eq:lem_reduction3}
 \end{align}

The next lemma further generalizes Lemma~\ref{lem:path_reduction} when there are $r$ intermediate vertices $v_{k_r}$ along the paths joining $v_i$ and $v_j$.

\begin{lemma}[Multiple intermediate vertices] \label{lem:multi_intermediate_vertex}
Consider an undirected matrix-weighted graph $G$. Let $r \in \{2,\ldots,n-1\}$ and define
\begin{align*}
{\rm ker}(\mc{Q}_{ij}^1) &= {\rm ker}(\m{A}_{ij}), \\
{\rm ker}(\mc{Q}_{ij}^r) &= \bigcap_{k_{r-1} = 1}^n \left( \bigcap_{k_{r-2} =1}^n \left(\ldots \bigcap_{k_1=1}^n \left( {\rm ker}(\mc{P}_{ik_1}) + {\rm ker}(\mc{P}_{k_1k_2})\right) + \ldots + {\rm ker}(\mc{P}_{k_{r-2}k_{r-1}}) \right) + {\rm ker}(\mc{P}_{k_{r-1}j}) \right)
\end{align*}
If for some $r\in \{1,\ldots,n-1\}$ we have
\begin{align} \label{eq:multi_vertex_condition}
    {\rm dim}\left({\rm ker}(\mc{Q}_{ij}^r)\right)=0,
\end{align}
then two vertices $v_i\ne v_j$ belong to a same cluster. Note that the immediate vertex can be $v_i$ or $v_j$.
\end{lemma}

\begin{proof}
For $r=1$, we only consider ${\rm ker}(\mc{P}_{ij}) = {\rm ker}(\m{A}_{ij})$. In this case, the immediate vertex can be $v_i$ or $v_j$. The case $r=2$ has been proved in Lemma~\ref{lem:path_reduction}. For $r>2$, the proof follows by applying Lemma~\ref{lem:path_reduction} several times and the observation that $\mc{Q}_{ij}^r$ imposes a constraint on $\m{x}_i - \m{x}_j$ corresponding to all paths joining $v_i,v_j$ passing immediate vertices $v_{k_1},\ldots,v_{k_{t-1}}$.
\end{proof}

\subsection{Warshall algorithm}
\label{subsec:warshall_alg}
In this subsection, we propose Warshall algorithm for matrix-weighted graphs. Let $G=(V,E,W)$ be an undirected matrix-weighted graph of $n$ vertices with the corresponding matrix-weighted adjacency matrix $\m{A}=[\m{A}_{ij}]$. Since using the condition~\eqref{eq:multi_vertex_condition} directly for assessing connectivity would be messy, we use matrices to record immediate computing steps of $\mc{Q}_{ij}^r$. The subspace sum and subspace intersection are performed by matrix operators ``$\wedge$'' and ``$\vee$'' on the corresponding matrices.

In matrix-weighted graph, for any two vertices $v_i$ and $v_j$, three possible scenarios may happen: (i) $v_i$ and $v_j$ belong to a same cluster, i.e., $v_i$ and $v_j$ are connected; (ii) $v_i$ and $v_j$ does not belong to a same cluster but there exists at least a path in $G$ joining $v_i$ and $v_j$; and (iii) there is no path joining $v_i$ and $v_j$. Since connectedness between two clusters of a matrix-weighted graph is an aggregated efforts between different paths in the graph, in addition to the connected state (case (i)) and the disconnected state (case (iii)), we define an \emph{undecided state} which is only determined after a certain algebraic condition was satisfied. 
For a positive semidefinite matrix $\m{A}_{ij} \in \mb{R}^{d\times d}$, the decision operator $\mc{D}$ is defined as follows:
\begin{align} \label{eq:decision_operator}
\mc{D}(\m{A}_{ij}) &= \left\lbrace \begin{array}{ll}
\m{I}_d, & \text{if } {\rm dim}(\m{A}_{ij}) = d,\\
\bm{\Theta}_d, & \text{if } \m{A}_{ij} = \bm{\Theta}_d,\\
\m{A}_{ij}, & \text{if } {\rm dim}(\m{A}_{ij}) < d,
\end{array} \right.
\end{align}
and for a block matrix $\m{A}=[\m{A}_{ij}]$,
\begin{align}
\mc{D}(\m{A}) &= [\mc{D}(\m{A}_{ij})].
\end{align}

Several properties of the decision operator are given in Lemma~\ref{lem:decision_operator}, whose proof can be found in~\ref{app:decision}.
\begin{lemma}[The ``$\mc{D}$'' operator] \label{lem:decision_operator}
Let $\m{A}_1,\m{A}_2\in\mb{R}^{d\times d}$ be symmetric positive semidefinite matrices, there holds
\begin{itemize}
    \item[i.]  ${\rm ker}(\mc{D}(\m{A}_1))={\rm ker}(\m{A}_1)$,
    \item[ii.] ${\rm ker}(\mc{D}(\m{A}_1 \vee \m{A}_2))={\rm ker}(\mc{D}(\m{A}_1) \vee \mc{D}(\m{A}_2))$,
    \item[iii.] ${\rm ker}(\mc{D}(\m{A}_1 \wedge \m{A}_2))={\rm ker}(\mc{D}(\m{A}_1) \wedge \mc{D}(\m{A}_2))$,
    \item[iv.] If $\m{A}_2$ is positive definite, ${\rm ker}(\mc{D}(\m{A}_1 \wedge \m{A}_2))={\rm ker}(\m{A}_1)$ and $\mc{D}(\m{A}_1 \vee \m{A}_2)=\m{I}_d$.
\end{itemize}
\end{lemma}

Let $\m{A}=[\m{A}_{ij}]$ and $\m{B}=[\m{B}_{ij}] \in \mb{R}^{dn \times dn}$ be block matrices, each submatrix is of size $d\times d$. The ``$\wedge$'' operator for two block matrices $\m{A}$ and $\m{B}$ is defined as $\m{C}= [\m{C}_{ij}] =\m{A} \wedge \m{B}$, where
\begin{align} \label{eq:c4_A1}
\m{C}_{ij} =\bigvee_{k=1}^n(\m{A}_{ik} \wedge \m{B}_{kj}),~\forall i,j=1,\ldots,n,
\end{align}
and the ``$\vee$'' operator for two block matrices $\m{A}$ and $\m{B}$, denoted as $\m{D}=[\m{D}_{ij}] = \m{A} \vee \m{B}$, satisfies
\begin{align} \label{eq:c4_A2}
\m{D}_{ij} = \m{A}_{ij} \vee \m{B}_{ij},~\forall i,j=1,\ldots,n.
\end{align}
The `$\wedge$' and `$\vee$' operators for block matrices are defined as  generalizations of the usual matrix multiplication and matrix addition.

We define the power of a block matrix $\m{A}=[\m{A}_{ij}]$ combined with the decision operator recursively as follows
\begin{subequations} \label{eq:matrix_power}
\begin{align}
\m{A}^0 &= \m{I}_{dn}, \\
\m{A}^1 &= \mc{D}(\m{A}),\\
\m{A}^k &= \mc{D}(\m{A}^{k-1} \wedge \m{A}),\, k \geq 2.
\end{align}
\end{subequations}

For an $n$-vertex matrix-weighted graph $G$, let
\begin{subequations} \label{eq:matrix_M}
\begin{align} 
\m{M}(G,0) &= \m{A}^0=\m{I}_{dn}, \label{eq:matrix_M_1}\\
\m{M}(G,k) &= \mc{D}\left(\m{M}(G,k-1) \vee \m{A}^k\right),\, k\ge 1. \label{eq:matrix_M_2}
\end{align}
\end{subequations}
The combinations of the operators ``$\vee$'', ``$\wedge$'' and ``$\mc{D}$'' in equation~\eqref{eq:matrix_M} is characterized in the following lemmas, whose proof can be found in \ref{app:monotone} and \ref{app:kernel2matrix}. 

\begin{lemma} \label{lem:kernelSpaces2matrix}
Consider an undirected matrix-weighted graph $G$ and consider the power matrices defined as in \eqref{eq:matrix_power}. For $r=1,\ldots,n-1$, let ${\rm ker}(\mc{R}_{ij}^t)$ denote the kernel space of all paths of length $t$ joining $v_i \ne v_j$, then ${\rm ker}(\mc{R}_{ij}^t) = {\rm ker}\left([\m{A}^t]_{ij}\right)$. If ${\rm dim}\left(\bigcap_{t=1}^{n-1}{\rm ker}(\mc{R}_{ij}^t)\right)=0$, then $v_i,v_j$ are connected.
\end{lemma}

\begin{lemma}[Monotonicity in connectivity] \label{lem:monotonicity_kernel}
Consider an undirected matrix-weighted graph ${G}$ which has the matrix-weighted adjacency matrix $\m{A} \in \mb{R}^{dn \times dn}$ with block matrices $\m{A}_{ij} \in \mb{R}^{d\times d}$. With the matrices $\m{M}(G,t),~t=1,\ldots,n-1,$ defined as in \eqref{eq:matrix_M}, we have
\begin{itemize}
    \item[i.]  ${\rm ker}([\m{M}(G,t)]_{ij}) = {\rm ker}\left(\left[\mc{D}\left(\bigvee_{i=0}^t\m{A}^k\right)\right]_{ij}\right)$, and
    \item[ii.] ${\rm ker}([\m{M}(G,t))]_{ij}\subseteq {\rm ker}([\m{M}(G,t-1)]_{ij})$, where $[\m{M}(G,t)]_{ij}$ denotes the $ij$-th block matrix in $\m{M}(G,t)$.
\end{itemize}
\end{lemma}

Algorithm~\ref{alg:walshall}, the Warshall algorithm,\footnote{We use the name Warshall algorithm as it uses an analogous boolean matrix multiplication as in the Warshall algorithm for topological graph \citep{Warshall1962}.} is proposed to determine connectedness and clustering of an undirected matrix-weighted graph $G$. The main result of this subsection is stated in the following theorem. 

\begin{algorithm}[t!]
\caption{Warshall algorithm \label{alg:walshall}}
\begin{algorithmic}[1]
\REQUIRE Matrix weighted graph $G=(V,E,W)$ with matrix-weighted adjacency matrix $\m{A}$ and terminal time $T_{\rm end}$
\ENSURE A partition $\mc{C}_G$ of vertices in $V$ into clusters 
\STATE $k \gets 0$ 
\STATE $\m{M} \gets \m{I}_{dn}$
\STATE $\m{A}_k \gets \mc{D}(\m{A})$

\REPEAT
   \STATE $\m{M} \gets \mc{D}(\m{M} \vee \m{A}_k)$
   \STATE $\m{A}_k \gets \mc{D}(\m{A}_k \wedge \m{A})$
   \STATE $k \gets k+1$
\UNTIL{$k\ge \min\{n-1,T_{\rm end}\}$ \OR $\m{M}==\m{1}_n \m{1}_n^\top \otimes \m{I}_d$}
\IF{$\m{M}\ne \m{1}_n\m{1}_n^\top\otimes\m{I}_d$}
   \STATE $\mc{C}_G(1) \gets \{\mc{C}_j=\{v_j\},j=1,\ldots,n\}$
   \STATE $i \gets 1$
   \REPEAT
      \STATE $j \gets i+1$
      \REPEAT
		\IF{$\m{M}_{ij}==\m{I}_d$}
		    \IF{$\mc{C}_i \in \mc{C}_G(i)$ \AND $\mc{C}_j \in \mc{C}_G(i)$}
			   \STATE $\mc{C}_i \gets \mc{C}_i \cup \mc{C}_j$
			   \STATE $\mc{C}_G(i+1) \gets \mc{C}_G(i) \setminus \mc{C}_j$
			\ENDIF
		\ENDIF
		\STATE $j \gets j+1$
	  \UNTIL{$j>n$}
	  \STATE $i \gets i+1$
   \UNTIL{$i\ge n$}
   \RETURN $\mc{C}_G(i)$
  \ELSE
  \RETURN $\mc{C}_G \gets \{V\}$
\ENDIF
\end{algorithmic}
\end{algorithm}

\begin{theorem}\label{thm:warshall}
Consider an undirected matrix-weighted graph $G=(V,E,W)$ and $T_{\rm end}>n$. The following claims on Algorithm~\ref{alg:walshall} hold.
\begin{itemize}
\item[i.] Two vertices $v_i,v_j \in V$ belong to the same cluster if there exists $r \le n-1$ such that the $ij$-th block matrix of $\m{M}(G,r)$ satisfies $[\m{M}(G,r)]_{ij}=\m{I}_d$.
\item[ii.] The graph $G$ is connected if $\m{M}(G,n-1)=\m{1}_{n}\m{1}_{n}^\top \otimes \m{I}_d$. If there is no path joining $v_i$ to $v_j$, then $[\m{M}(G,n-1)]_{ij}=[\m{M}(G,n-1)]_{ji}=\bm{\Theta}_d$. 
\end{itemize}
\end{theorem}
\begin{proof}
    i. Since $[\m{M}(G,r)]_{ij} = \mc{D}\left( \bigvee_{t=0}^{r}\left( [\m{A}^t]_{ij} \right) \right)$, for $i\ne j$, ${\rm ker}([\m{M}(G,r)]_{ij})=\bigcap_{t=1}^{r}{\rm ker}\left([\m{A}^t]_{ij}\right)$. Based on Lemma~\ref{lem:kernelSpaces2matrix}, if $[\m{M}(G,r)]_{ij}=\m{I}_d$ then $v_i,v_j$ belong to a same cluster.

    ii. Based on (i), it is clear that $G$ is connected if all $v_i,v_j$ belong to a same cluster, or equivalently, $\m{M}(G,n-1)=\m{1}_n\m{1}_n^\top\otimes\m{I}_d$. Trivially, if $v_i$ and $v_j$ is topologically disconnected, there exists no path joining them. Thus, $[\m{M}(G,n-1)]_{ij}=\bm{\Theta}_d$. 
\end{proof}
\begin{figure}[ht]
\centering
\includegraphics[scale=1]{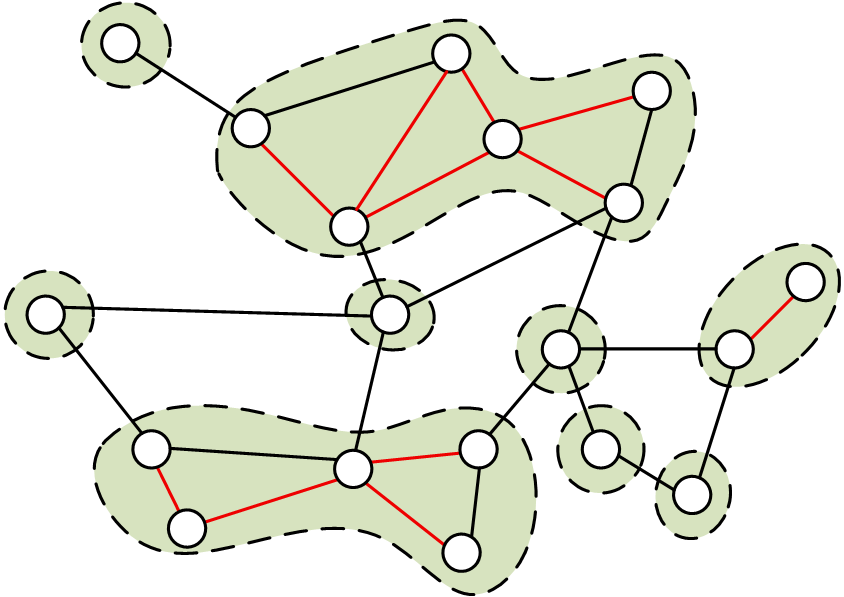}
\caption{If $T_{\rm end}=1$, each cluster in the output partition of Algorithm~\ref{alg:walshall} may be a positive tree or a single vertex. \label{fig:one_step}}
\end{figure}

\begin{figure}[ht]
\centering
\subfloat[]{\includegraphics[width=0.3 \linewidth]{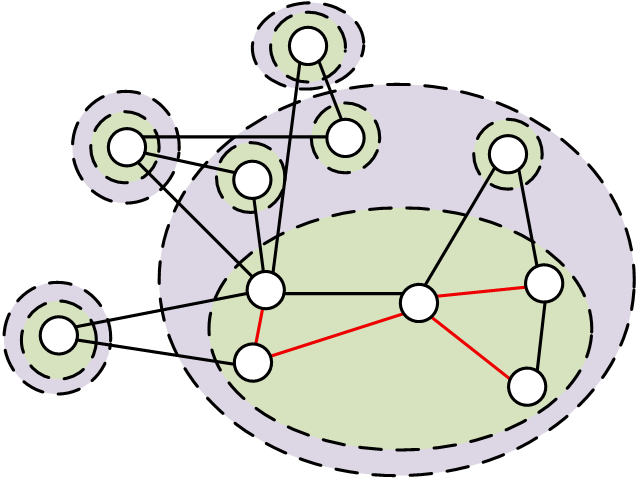}}
\hfill
\subfloat[]{\includegraphics[width=0.3 \linewidth]{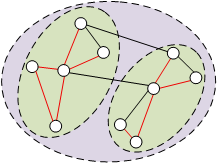}}
\hfill
\subfloat[]{\includegraphics[width=0.33 \linewidth]{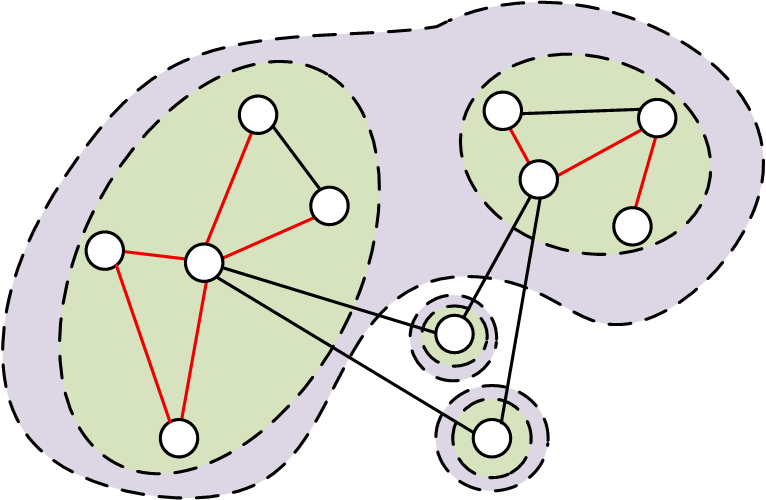}}
\\
\subfloat[]{\includegraphics[width=0.65 \linewidth]{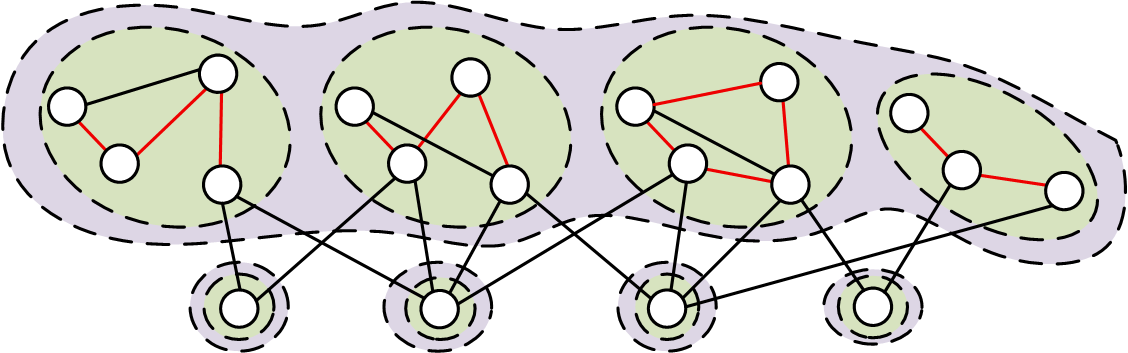}}
\caption{If $T_{\rm end}=2$, there are many possible configurations for each cluster in the output partition of Algorithm~\ref{alg:walshall}. \label{fig:two_step}}
\end{figure}

\begin{remark}[Output partition with early termination]
We consider the case $T_{\rm end}<n$, and the connectivity condition in  \eqref{lem:multi_intermediate_vertex} is tested until $T_{\rm end}$. 
\begin{itemize}
\item[i] If $T_{\rm end}=1$, each cluster in the output partition has a spanning positive tree or contains a single vertex.  Fig.~\ref{fig:one_step} depicts several possible cluster configurations. Each red line depicts an edge with positive definite matrix weight and each black line depicts an edge with positive semidefinite matrix weight. Each cluster that can be identified with one-step termination is encircled by a dashed line and colored in green. This fact can be proved by observing that in the first step, each $v_i$ knows $v_j \in \mc{N}_i$ belongs to the same cluster if and only if $\m{A}_{ij}\succ 0$. Let $v_i \sim v_j$ denote $\m{A}_{ij}\succ 0$. Lines 9--27 groups vertices into cluster. Consider $v_1$. All vertices $v_{i_1} \sim v_1$ are included into $\mc{C}_1$. These vertices are then again checked, and all vertices $v_{i_2} \sim v_{i_1}$ are also included into $\mc{C}_1$. As the algorithm continues including vertices $\{v \notin \mc{C}_1 \mid \exists w \in \mc{C}_1,~v \sim w \}$, there is a path joining $v_1$ to every other vertices in $\mc{C}_1$ having all positive definite edge weights.
\item[ii] For a set of edge $E' \subseteq E$, the weight of $E'$ is defined as the sum of all matrix weights in $E'$. If $T_{\rm end}=2$, each cluster in the output partition may contain a single vertex, or a positive tree, or a positive tree $\mc{T}'$ and vertices $v$ such that the sum of the matrix weights between $v$ and $V(\mc{T}')$ is positive definite, or positive trees $\mc{T}'_k$ joining to each other by a set of edges with positive definite weight or via 1-hop intermediate vertex(ices). As depicted in Fig.~\ref{fig:two_step}, with $T_{\rm end}=2$, various possible final clustering configurations may exist. Each cluster that can be identified after two-step termination is encircled by a dashed line and colored in purple.
\item[iii] For each vertex $v \in V$, it is helpful to think of each iteration $k$ as one step expansion from the graph centered at $v$ (with vertex set $\{w \in G \mid {\rm dist}(v,w) \le k-1\}$) to vertices which are $k$-neighbor separated from $v$. The information are aggregated from every vertices (upon the terminal condition in line 8 of Algorithm~\ref{alg:walshall} holds) to provide an approximation of clusters in the graph up to the available information.
\end{itemize} 
\end{remark}

\begin{remark}[Computational complexity of Algorithm~\ref{alg:walshall}]\label{remark:complexity_Warshall}
We provide a worst-case computational complexity for the Algorithm~\ref{alg:walshall} as follows.
\begin{itemize}
\item[i.] Determine the graph's topology using a Depth-First Search algorithm is $O(|V|+|E|)$ \citep{Rosen2019discrete}. In the worst-case, $G$ is a complete graph and the computational complexity is upper bounded by $O(n^2)$.

\item[ii.] Due to symmetry of matrices $\m{A}^{r},~r=0,\ldots,t-1$, computing $\m{A}^{t-1} \wedge \m{A}$ requires computing $\frac{1}{2}n(n+1)$ submatrices $[\m{A}^t]_{ij}$, with $1<i<j\le n$ according to \eqref{eq:c4_A1}. To compute each submatrix in \eqref{eq:c4_A1}, we need $n$ parallel matrix addition $[\m{A}^{t-1}]_{ik}\wedge \m{A}_{kj}$ \eqref{eq:parallel_sum} and $n-1$ series matrix addition.\footnote{Since series matrix addition requires finding a set of independent eigenvectors, $n-1$ series matrix addition can be perform only one time, after $n$ parallel matrix addition have been completed.} Thus, computing $[\m{A}^t]_{ij}$ is of $O(nd^3)$, and $\m{A}^{t}$ is of $O(n^3d^3)$.

\item[iii.] To determine $\m{M}(G,t)$, the symmetry also reduces the computation burden. As $[\m{M}(G,t) \vee \m{A}^t]_{ii}=\m{I}_d,~\forall i$, we need to compute the ``$\vee$'' operator $\frac{(n-1)n}{2}$ times to determine $[\m{M}(G,t-1) \vee \m{A}^t]_{ij}$, $1<i<j\le n$. The computations from the previous steps can be reused here, with the cost of storing eigenvectors and eigenvalues decomposition of $[\m{M}(G,t-1)]_{ij}$ and $[\m{A}^t]_{ij}$. Thus, a computational complexity of $O(d^3)$ is required for determining a set of linearly independent eigenvectors from those in $[\m{M}(G,t-1)]_{ij}$ and $[\m{A}^t]_{ij}$. If the calculations from previous steps are not saved in the memory, the computational cost is tripled, as we need to find the set of eigenvectors and eigenvalues of each matrix $[\m{M}(G,t-1)]_{ij}$ and $[\m{A}^t]_{ij}$ again before determining the intersection of them for $[\m{M}(G,t-1) \vee \m{A}^t]_{ij}$. 

After this step, the decision operator $\mc{D}$ is immediately applied $\mc{D}([\m{M}(G,t-1) \vee \m{A}^t]_{ij})$, and it is reasonable to assume that the set of linearly independent eigenvectors of $[\m{M}(G,t-1) \vee \m{A}^t]_{ij}$ is still available. Thus, $\mc{D}([\m{M}(G,t-1) \vee \m{A}^t]_{ij})$ adds $O(d)$ to the computational complexity. 

To sum up, determining $\m{M}(G,t)$ is of $(O(d^3)+O(d))\frac{1}{2}n(n-1)=O(d^3n^2)$.
\item[iv.] In the worst-case, we need to compute until $t=n-1$ to obtain $\m{M}(G,n-1)$. The total computational cost is thus upper bounded by $O(n^2) + (O(n^3d^3)+O(n^2d^3))(n-2) = O(n^4d^3)$.
\end{itemize}
\end{remark}

\begin{remark}
It is helpful to compare Warshall algorithms for topological graph and matrix-weighted graphs. Let $G=(V,E,W)$ be an undirected matrix-weighted graph. Defining the Boolean adjacency matrix $\m{A}_s =[a_{ij}^s] \in \mb{R}^{n \times n}$, $a_{ij}^s=1$ if $\m{A}_{ij}\ne \bm{\Theta}_d$ and $a_{ij}=0$, otherwise. It is well known that $G$ is topologically connected if and only if 
\begin{align} \label{eq:scalar_Warshall}
\m{M}_s(G,n-1)=\m{I}_n + \m{A}_s+ \m{A}^2_s + \ldots + \m{A}^{n-1}_s
\end{align}
is a positive matrix \citep{Rosen2019discrete,Warshall1962}. Otherwise, the graph contains disconnected components that can be identified from the zero patterns of $\m{M}_s$. This property holds due to the fact that $\m{A}_s$ is a nonnegative matrix. 
However, nonnegativity does not hold for a matrix-weighted adjacency matrix $\m{A} \in \mb{R}^{dn \times dn}$. The iteration for determining $\m{M}(G,n-1)$ in the Warshall algorithm mimics the summation \eqref{eq:scalar_Warshall} by replacing the sum and multiplication with the ``$\vee$'' and ``$\wedge$'' operators, respectively. The lack of nonnegativity is partly responsible for the greedy nature of Algorithm~\ref{alg:walshall}.
\end{remark}

\subsection{Distributed Warshall algorithms}
\label{subsec:Distributed_Warshall}
In this subsection, we propose a distributed version of the Warshall algorithm~\ref{alg:walshall}. Consider a network of $n$ agents equipped with local computational power, memory and networking capability. The interactions of the agent is characterized by a connected undirected graph $(V,E)$. We would like to determine whether a given set of positive semidefinite matrix weights ${W}$ assigned to each edge of $(V,E)$ would result in a connected matrix-weighted graph $G=(V,E,W)$ in a distributed manner. 

Suppose that each agent $i$ knows $\m{A}_{ij}=\m{A}_{ji} \in \mb{R}^{d \times d}$, $\forall j \in \mc{N}_i$. Agent $i$ has state matrices $\m{X}_i(t), \m{Y}_i(t) \in \mb{R}^{d \times d}$ with initial condition $\m{X}_i(0)=\m{X}_i^\top(0)$, $\m{Y}_i(0)= \m{Y}_i^\top(0)\in \mb{R}^{d \times d}$. Consider the following recursive relation
\begin{subequations} \label{eq:d_Warshall}
\begin{align}
\m{Y}_i(t+1) &=  \mc{D}\left(\m{X}_i(t) \vee \m{Y}_i(t) \right), \label{eq:d_Warshall1} \\
\m{X}_i(t+1) &= \mc{D}\left(\bigvee_{j = 1}^n(\m{A}_{ij} \wedge \m{X}_j(t))\right), \label{eq:d_Warshall2}
\end{align}
\end{subequations}
for $i=1,\ldots,n,$ and $t=0, 1, \ldots, n-1$. Clearly, \eqref{eq:d_Warshall2} requires agent $i$ to obtain $\m{X}_j(t)$ and broadcast $\m{X}_i(t)$ to its neighboring agents at each iteration $t$. The algorithm \eqref{eq:d_Warshall1} replaces the usual matrix multiplications and matrix summations in the matrix-weighted consensus algorithm \citep{Trinh2018matrix} with two operators ``$\wedge$'' and ``$\vee$'', respectively.

Defining $\mb{X}(t) = [\m{X}_1^\top(t),\ldots,\m{X}_n^\top(t)]^\top\in \mb{R}^{dn \times d}, \mb{Y}(t) = [\m{Y}_1^\top(t),\ldots,\m{Y}_n^\top(t)]^\top \in \mb{R}^{dn \times d}$, we can express \eqref{eq:d_Warshall} in the following matrix form
\begin{subequations} \label{eq:d_Warshall_m}
\begin{align}
\mb{Y}(t+1) &= \mc{D}(\mb{X}(t) \vee \mb{Y}(t)),\label{eq:d_Warshall_m1}\\
\mb{X}(t+1) &=  \mc{D}(\m{A} \wedge \mb{X}(t)), \label{eq:d_Warshall_m2}
\end{align}
\end{subequations}
for $t=0,1,\ldots,n-1$ and initial conditions $\mb{X}(0)$, $\mb{Y}(0)$, where the matrix-vector ``$\wedge$'' product \eqref{eq:d_Warshall_m2} is defined as in \eqref{eq:d_Warshall2}.
 
Equations~\eqref{eq:d_Warshall_m1} and \eqref{eq:d_Warshall_m2} can be expanded as follows
\begin{align}
\mb{Y}(t+1) &= \mc{D}\left(\mb{X}(t) \vee \mc{D}\left(\mb{X}(t-1) \vee \cdots \vee \mb{X}(0)\right)\right) \label{eq:iter_2} \\
\mb{X}(t+1) &= \mc{D}(\m{A} \wedge \mb{X}(t)), \nonumber\\
&= \mc{D}\left(\m{A} \wedge \mc{D}\left( \m{A} \wedge \mb{X}(t-1) \right)\right) \nonumber \\
&\qquad \vdots \nonumber\\
&= \mc{D}\left( \m{A} \wedge \mc{D}(\cdots \wedge \mc{D}(\m{A} \wedge \mb{X}(0))\right), \label{eq:iter_1}
\end{align}
As the recursion terminates at $t=n-1$, we obtain $\mb{Y}(n)$. Based on the equations \eqref{eq:iter_1}--\eqref{eq:iter_2},  Algorithm~\ref{alg:distributed_walshall}, the distributed Warshall algorithm, is proposed for matrix-weighted graphs. The following theorem is about the convergence of Algorithm~\ref{alg:distributed_walshall}.

\begin{algorithm}[t!]
\caption{Distributed Warshall algorithm \label{alg:distributed_walshall}}
\begin{algorithmic}[1]
\REQUIRE Each agent $i = 1,\ldots,n$ knows $\m{A}_{ij},~j\in\mc{N}_i$
\ENSURE The cluster $\mc{C}_i$ that contains $v_i$
\STATE $t \gets 0$ 
\STATE Agent $i$ initializes $2n$ matrix variables: \\ 
$\m{X}_{i}^k(0)\gets \mc{D}(\m{A}_{ik})$, $\forall k \in \{1,\ldots,n\}$ \\
$\m{Y}_{i}^i(0)\gets \m{I}_d$, $\m{Y}_{i}^k(0)\gets \bm{\Theta}_d$, $\forall k \in \{1,\ldots,n\} \setminus \{i\}$

\REPEAT
   \STATE Agent $i$ updates it state variables: For $k = 1,\ldots,n$,\\
	$\m{Y}_i^k(t+1) \gets \mc{D}\left(\m{X}_i^k(t) \vee \m{Y}_i^k(t)\right),$\\
	$\m{X}_i^k(t+1) \gets \mc{D}\left(\bigvee_{j \in \mc{N}_i}\left(\m{A}_{ij} \wedge \m{X}_j^k(t) \right) \right),$\\
   \STATE $t \gets t+1$
\UNTIL{$t\ge n-1$}

\STATE Agent $i$ determines the cluster it belongs to: $\mc{C}_i \gets \{v_j|~ {\rm rank}(\m{Y}_i^j(n-1)) = d\}$
\RETURN $\mc{C}_i$
\end{algorithmic}
\end{algorithm}

\begin{theorem}\label{thm:distributed_Warshall}
After the Algorithm~\ref{alg:distributed_walshall} terminates, the matrix-weighted graph $G$ is connected if $\mc{C}_i = V$, $\forall i=1,\ldots,n$.
\end{theorem}
\begin{proof}
Let 
\begin{align*}
\mb{X}^k(t) &= [\m{X}_1^k(t),\ldots,\m{X}_n^k(t)]^\top \in \mb{R}^{dn \times d},~
\mathfrak{X}(t) = [\mb{X}^1(t),\ldots,\mb{X}^n(t)]\in \mb{R}^{dn \times dn}, \\
\mb{Y}^k(t) &= [\m{Y}_1^k(t),\ldots,\m{Y}_n^k(t)]^\top \in \mb{R}^{dn \times d},
~ \mathfrak{Y}(t) = [\mb{Y}^1(t),\ldots,\mb{Y}^n(t)]\in \mb{R}^{dn \times dn}.
\end{align*}
Since $\mathfrak{X}(0)=\m{A}$, $\mathfrak{Y}(0)=\m{I}_{dn}=\m{A}^0$, we can express the $n$ parallel iterations in Algorithm~\ref{alg:distributed_walshall} as follows
\begin{align}
\mathfrak{X}(t) &= \mc{D}(\m{A} \wedge \mathfrak{X}(t-1)) \nonumber\\
&= \mc{D}\left(\m{A} \wedge \mc{D}( \m{A} \wedge (\cdots \wedge \mc{D}(\m{A} \wedge \mathfrak{X}(0))))\right) \nonumber\\
&\qquad \vdots \nonumber\\
&= \m{A}^{t+1}, 
\end{align}
\begin{align}
\mathfrak{Y}(t+1) &= \mc{D}(\mathfrak{X}(t) \vee \mathfrak{Y}(t)) = \mc{D}\left( \m{A}^{t+1} \vee \mathfrak{Y}(t) \right) \nonumber\\
&\qquad \vdots \nonumber\\
&= \mc{D}\left( \m{A}^{t+1}  \vee \mc{D}\Big( \m{A}^{t-1} \vee \ldots \vee \mc{D}(\m{A} \vee \mathfrak{Y}(0)) \Big)\right) \nonumber \\
&= \m{M}(G,t+1),
\end{align}
for $t=0,1,\ldots,n-2$. Thus, after the repeat loop ends, agent $i$ has row blocks $d(i-1)+1$ to $di$ of the matrix $\m{M}(G,n-1)$ saved in $n$  variables $\m{Y}^k_i(n-1)$, $k=1,\ldots,n$. The agent uses this information to determine $\mc{C}_i$. 

Based on Theorem~\ref{thm:warshall}~ii, if $\mc{C}_i$ contains all vertices of $G$, then $G$ is connected.
\end{proof}

\begin{remark}[Computational complexity of Algorithm~\ref{alg:distributed_walshall}] We estimate the computational complexity of the distributed algorithm \ref{alg:distributed_walshall} for each individual agent. The computational complexity of each iteration $t$ in \eqref{eq:d_Warshall} is $O(|\mc{N}_i|d^3)$. Since each agent executes $n$ parallel independent instances of \eqref{eq:d_Warshall}, one for each $k = 1,\ldots,n$, the per-iteration computational complexity (lines 3--6 of algorithm \ref{alg:distributed_walshall}) is upper bounded by $O(|\mc{N}_i|nd^3)$. The algorithm terminates after at most $t=n-1$ iterations yielding a worst-case total computational complexity of agent $i$ being $O(|\mc{N}_i|n^2d^3)$.

Furthermore, based on the handshaking lemma~\citep{Rosen2019discrete}, $\sum_{i=1}^n|\mc{N}_i| = 2 |E| = 2m$, and thus the worst-case average computational complexity of each agent is $O(mnd^3)$. For complete graph, $m=O(n^2)$, and thus, the computational complexity in the worst case is $O(n^3d^3)$. The overall $n$-agent system's computational complexity is of $O(n^4d^3)$ and is shared to $n$ agents.  
\end{remark}

\section{Numerical examples}
\label{sec:examples}
In this section, we consider several examples to illustrate the Warshall algorithm. 
\subsection{Example 1: A one-hundred-vertex graph}
\label{subsec:eg1}
\begin{figure*}
\centering
\subfloat[]{%
\input{MWG100.tex}
}
\qquad 
\subfloat[]{%
\input{MWG100b.tex}
}
\caption{Matrix-weighted graphs of 100 vertices with four types of matrix weights in Example~1a (left) and~1b (right).}
\label{fig:eg1}
\end{figure*}

\subsubsection{Example 1a: Connected graph}
Consider a 100-vertex matrix-weighted graph $G=(V,E,W)$ as depicted in Fig.~\ref{fig:eg1} a. The edges of $G$ are randomly generated so that an edge joining each vertex pair $v_i,v_j$ exists with probability $p=0.08$. Among the edges, there are four types of matrix weights 
\begin{align*}
\m{W}_{1} = \begin{bmatrix}
1 & 0 \\ 0 & 0
\end{bmatrix},\m{W}_2= \begin{bmatrix}
0 & 0 \\ 0 & 1
\end{bmatrix},\m{W}_3= \begin{bmatrix}
1 & 1 \\ 1 & 1
\end{bmatrix},\m{W}_4= \begin{bmatrix}
1 & 0 \\ 0 & 1
\end{bmatrix},
\end{align*}
with the corresponding probabilities $p_1 = p_2 = p_3 = \frac{0.8}{3}$ and $p_4 = 0.2$.

In our simulation test, the graph $G$ in Fig.~\ref{fig:eg1}a was generated with the above described statistics. Specifically, we have $n=100$, $m=434$ (with 87 positive definite edges), and $d=2$. The matrix-weighted Laplacian of $G$ satisfies ${\rm rank}(\m{L})=198 = dn-d$, i.e., $G$ is connected.

Using Algorithm~\ref{alg:walshall}, we found that the algorithm terminates at $k = 6<n-1 = 99$ and $\m{M}(G,6)=\m{1}_{100}\m{1}_{100}^\top\otimes\m{I}_2$. The numerical result verifies that the matrix-weighted graph $G$ is connected. 

It is worth noting that $k$, the smallest number of iterations at which Algorithm~\ref{alg:walshall} terminates and declares $G$ is connected, provides an upper bound of the diameter of the matrix-weighted graph. Each pair of vertices in $G$ are separated by a collection of paths with length at most 6.

\subsubsection{Example 1b: Clustered graph}

Next, we consider a 100-vertex graph randomly generated with $p=0.05$ and the same parameters $p_1,\ldots,p_4$ as in the previous example (Fig.~\ref{fig:eg1}b). Since the matrix-weighted Laplacian has ${\rm rank}(\m{L})=189$, the graph is clustering. We use Warshall algorithm to estimate a partition of the graph. 

Observe that if $G$ is not connected, the algorithm has to proceed precisely $n$ iterations (corresponding to the worst case computational complexity). 
To mitigate this limitation, a possible approach is modifying the algorithm to terminate once no change in the kernels of $[\m{M}(G,k+1)]_{ij}$ and $[\m{M}(G,k)]_{ij}$, $\forall i,j \in \m{M}$, can be detected. A better approach comes from the fact that social networks usually have small diameter, thus, it is unlikely to find new members of a cluster by searching deep and distanced connections. 

From these observations, we terminate the algorithm after $k=7$ steps, and obtain a partition of 10 clusters. The partition contains a large cluster containing 91 vertices and 9 small clusters of only 1 vertex. The output is heuristic, however, provides useful information on the connectivity inside the graph. From the output partition, we can form a reduced 10-vertex matrix-weighted graph to possibly reduce number of clusters.

\subsection{Example 2: A topologically connected matrix-weighted graph which is not connected}
\label{subsec:eg2}
In this example, a matrix-weighted graph of five vertices is considered ($n=5, d= 2$). The graph is depicted in Fig.~\ref{fig:graph-2} with the following matrix weights 
\begin{align*}
\m{A}_{12} &= \m{A}_{24}= \begin{bmatrix}
1 & 0\\0 &0
\end{bmatrix},~
\m{A}_{23} = \m{A}_{25} = \begin{bmatrix}
1 & -1\\-1 &1
\end{bmatrix},~\m{A}_{13}= \m{A}_{34}= \m{A}_{35} = \begin{bmatrix}
0 & 0\\0 &1
\end{bmatrix}.
\end{align*}

Note that $G$ has no positive definite edge. The rank of the matrix-weighted Laplacian is smaller than $8=dn-d$, which implies that the matrix-weighted graph is not connected.

\begin{figure}[t!]
\centering
{\resizebox{4.5cm}{!}{\begin{tikzpicture}[
roundnode/.style={circle, draw=black, thick, minimum size=3.5mm,inner sep= 0.25mm},
squarednode/.style={rectangle, draw=black, thick, minimum size=3.5mm,inner sep= 0.25mm},
]
    \node[roundnode] (u1) at (-0.5,-1.0) {$v_2$}; %
    \node[roundnode] (u2) at (-0.5,-2) {$v_3$};%
    \node[roundnode] (u3) at (-2.5,-1.5) {$v_1$};%
    \node[roundnode] (u4) at (1.5,-2)  {$v_4$};%
    \node[roundnode] (u5) at (1.5,-1.0){$v_5$};%
    \draw[-, very thick] (u5)--(u1);
    \draw[-, very thick] (u1)--(u2);
    \draw[-, very thick] (u2)--(u3);
    \draw[-, very thick] (u1)--(u3);
    \draw[-, very thick] (u2)--(u5);
    \draw[-, very thick] (u2)--(u4);
    \draw[-, very thick] (u1)--(u4);
\end{tikzpicture}}}
\caption{The matrix-weighted graph considered in Example~2}
\label{fig:graph-2}
\end{figure}
After using Algorithm~\ref{alg:walshall}, we obtain the matrix $\m{M}(G,4)$ as follows
\begin{align*} 
\m{M}(G,4) = \left[\begin{array}{cc|cc|cc|cc|cc}
1  & 0   &  0.95 &  0 &  0 &  0  &  1  &  0  &  0  &  0 \\
0    & 1   &  0 & 0&  0 &  1.125&  0    & 1 &  0  &  0.7179 \\
\hline
0.95 & 0 &    1 & 0&  0.7 & -0.7 &  0.95 &  0 & 0.7 &  -0.7 \\
0 &    0 &    0 & 1& -0.7 &  0.7 &  0 &   0 &   -0.7 &  0.7 \\
\hline
0 &  0    &0.7  & -0.7 &  1 &  0 & 0 &  0 & 1 &  0 \\
0 &  1.125& -0.7 &0.7 & 0  &  1 &  0 &  1.125 &  0  &  1 \\
\hline
1 &  0 &  0.95 &  0 & 0 & 0 & 1 & 0 &  0 &  0 \\
0 &  1 &  0   &  0 & 0 & 1.125 & 0  & 1 & 0 &   0.7179\\
\hline
0 &  0 & 0.7  & -0.7 & 1 & 0 & 0 &  0 &  1 & 0 \\
0 &  0.7179  & -0.7 &   0.7 & 0  &  1 &  0 &  0.7179 & 0  &  1
\end{array}\right].
\end{align*}

The pattern of identity submatrices matrix $[\m{M}(G,4)]_{ij} = \m{I}_2$ suggests that $G$ has three clusters $\mc{C}_1=\{v_1,v_4\}$, $\mc{C}_2=\{v_2\}$ and $\mc{C}_3=\{v_3,v_5\}$. It is interesting that two vertices $v_1$ and $v_4$ does not have any positive path joining them, and are not adjacent to each other, are actually lying on a same cluster.

We can use condition~\eqref{eq:kernel_Sij} to verify that $v_1$ and $v_4$ belong to a same cluster. The paths in $\mc{S}_{14}$ and their kernels are listed as follows: $\mc{P}_1 = v_1v_2v_4$, ${\rm ker}(\mc{P}_1) = {\rm im}\left(\begin{bmatrix}
0 \\ 1
\end{bmatrix}\right)$,~ $\mc{P}_2 = v_1v_3v_4$, ${\rm ker}(\mc{P}_2) = {\rm im}\left(\begin{bmatrix}
1 \\ 0
\end{bmatrix}\right)$,~ $\mc{P}_3 = v_1v_2v_3v_4$, ${\rm ker}(\mc{P}_3) = \mb{R}^2$, $\mc{P}_4 = v_1v_3v_2v_4$, ${\rm ker}(\mc{P}_4) = \mb{R}^2$, $\mc{P}_5 = v_1v_2v_5v_3v_4$, ${\rm ker}(\mc{P}_5) = \mb{R}^2$, $\mc{P}_6 = v_1v_3v_5v_2v_4$, ${\rm ker}(\mc{P}_6) = \mb{R}^2$. Therefore, we can compute ${\rm ker}(\mc{S}_{14}) = \bigcap_{k=1}^6 \mc{P}_k = \{\m{0}_2\}$ and ${\rm dim}({\rm ker}(\mc{S}_{14}))=0$. 

It is also observe that all submatrices $[\m{M}(G,4)]_{ij}$ are non-zero in this example, which suggests that the topological graph of $G$ is connected. However, this is actually not generally true that a connected matrix-weighted graph of $n$ vertices has all nonzero matrix block $[\m{M}(G,n-1)]_{ij}$. The next example illustrates this fact.

\subsection{\texorpdfstring{Example 3: Zero submatrix in $\m{M}(G,n-1)$ does not imply topological disconnectedness}{Example 3: Zero submatrix in M(G,n-1) does not imply topological disconnectedness}}
\begin{figure}[h!]
\centering
{\resizebox{3.35cm}{!}{\begin{tikzpicture}[
roundnode/.style={circle, draw=black, thick, minimum size=3.5mm,inner sep= 0.25mm},
squarednode/.style={rectangle, draw=black, thick, minimum size=3.5mm,inner sep= 0.25mm},
]
    \node[roundnode] (u1) at (0,0) {$v_1$}; %
    \node[roundnode] (u2) at (1,1) {$v_2$};%
    \node[roundnode] (u3) at (2,0) {$v_3$};%
    \node[roundnode] (u4) at (3,1)  {$v_4$};%
    \draw[-, very thick] (u1)--(u2)--(u3)--(u4);
\end{tikzpicture}}}
\caption{The matrix-weighted graph considered in Example~3}
\label{fig:graph-3}
\end{figure}

We consider a line graph of four vertices $G(V,E,W)$ in Fig.~\ref{fig:graph-3} with 
\begin{align*}
    \m{A}_{12} = \begin{bmatrix}
        0 & 0\\
        0 & 1
    \end{bmatrix}, \m{A}_{23} = \begin{bmatrix}
        1 & 0\\
        0 & 0
    \end{bmatrix}, \m{A}_{34} = \begin{bmatrix}
        1 & 0.5\\
        0.5 & 1
    \end{bmatrix}.
\end{align*}
Clearly, there is only one path from $v_1$ to $v_4$, which is $\mc{P} = v_1v_2v_3v_4$. It is not hard to verify that
${\rm ker}(\m{A}_{12})={\rm im}\left( \begin{bmatrix}
    1 \\ 0
\end{bmatrix}\right)$, ${\rm ker}(\m{A}_{23})={\rm im}\left( \begin{bmatrix}
    0 \\ 1
\end{bmatrix}\right)$, and ${\rm ker}(\m{A}_{34})=\{\m{0}_2\}$ and thus 
 ${\rm ker}(\mc{P}) = {\rm ker}(\m{A}_{12}) \cup {\rm ker}(\m{A}_{23}) \cup {\rm ker}(\m{A}_{34}) = \mb{R}^2.$
Thus, the matrix block $[\m{M}(G,3)]_{14}=\Theta_d$ while the graph is topologically connected. Indeed, the computation gives
\begin{align*}
\m{M}(G,3) = \left[\begin{array}{cc|cc|cc|cc}
1  &  0   &   0  &   0 &    0    &    0   & 0  &    0\\
0  &    1   &     0    &    0.75 &    0    &    0   & 0  &    0\\
\hline
0  &  0   &     1  &    0    &  0.9208    &    0   & 0.3  &   0\\
0  & 0.75   &     0    &    1    &    0    &    0   & 0  &    0\\
\hline
0  &    0   &     0.9208    &    0    &    1    &   0   & 1  &  0\\
0    &    0   &     0    &    0    &    0    &    1   & 0&    1\\
\hline
0    &    0   &     0.3  &    0    &    1    & 0    & 1  &    0\\
0    &    0   &     0    &    0    &    0  & 1    &0 & 1   \\
\end{array}\right],
\end{align*}
which verifies the computation based on path kernels. 

\section{Conclusion}
\label{sec:conclusion}
In this paper, the Warshall algorithm and distributed Warshall algorithm were proposed for matrix-weighted graphs. The algorithm simultaneously determines connectivity between every pair of distinct vertices in a matrix-weighted graph. The advantages of the Warshall algorithm include its ability to classify vertices into distinct clusters and its amenability to distributed implementation.

As an outlook, the matrix operators and heuristic approach in this paper seems to be peculiarly applicable for matrix-weighted graphs. As far as we concern, this approach for matrix-weighted graphs has not been adequately considered in the literature. Finally, from an application perspective, the proposed algorithms may be useful for fast detection and recovery of failed connection in a matrix-weighted graph.

\section*{Acknowledgment}

A preliminary version of this paper is available at \url{https://arxiv.org/pdf/2510.18260}. The Python code of the Warshall algorithm (Algorithm \ref{alg:walshall}) can be found at \url{https://github.com/
minhtrinh90/WarshallAlgorithmMWG.git}. ChatGPT was used for grammar checking and for suggestions on minor technical aspects of this paper. The authors are solely responsible for the content and conclusions of the paper.

\bibliographystyle{elsarticle-harv}
\bibliography{cas-refs}

\appendix
\numberwithin{equation}{section}

\section{Proof of Lemma~\ref{lem:parallel_sum}}
\label{app:parallel_sum}
i. For brevity, denote $\m{A}_s=\m{A}_1+\m{A}_2$. From the definition, $\m{A}_1\wedge\m{A}_2 = \m{A}_1\m{A}_s^\dagger\m{A}_2$, we have
\begin{align}
\m{A}_1 \wedge\m{A}_2 &= (\m{A}_s-\m{A}_2)\m{A}_s^\dagger(\m{A}_s-\m{A}_1) \nonumber\\
&=\m{A}_s - \m{A}_2 \m{A}_s^\dagger\m{A}_s - \m{A}_s\m{A}_s^\dagger \m{A}_1 +\m{A}_2\m{A}_s^\dagger\m{A}_1 \label{eq:A1}\\ 
&= \m{A}_2[\m{I}_d-\m{A}_s\m{A}_s^\dagger] + [\m{I}_d-\m{A}_s\m{A}_s^\dagger]\m{A}_1+\m{A}_2\m{A}_s^\dagger\m{A}_1, \label{eq:A2}
\end{align}
where we have used the identity $\m{A}_s\m{A}_s^\dagger\m{A}_s=\m{A}_s$ of the generalized inverse in equation \eqref{eq:A1}. It follows from \eqref{eq:A2} that $\m{A}_1\wedge\m{A}_2 - \m{A}_2\wedge\m{A}_1= \m{A}_2[\m{I}_d-\m{A}_s\m{A}_s^\dagger] + [\m{I}_d-\m{A}_s\m{A}_s^\dagger]\m{A}_1.$ Let $\m{x}$ be an arbitrary vector in $\mb{R}^d$, we have:
\begin{itemize}
\item  ${\rm ker}(\m{I}_d - \m{A}_s\m{A}_s^\dagger)={\rm im}(\m{A}_1+\m{A}_2) \subseteq {\rm im}(\m{A}_1)$, and thus $\m{A}_1\m{x} \in {\rm im}(\m{A}_1) \subseteq {\rm ker}(\m{I}_d - \m{A}_s\m{A}_s^\dagger)$, and $[\m{I}_d-\m{A}_s\m{A}_s^\dagger]\m{A}_1\m{x} = \m{0}_d$.
\item $(\m{I}_d-\m{A}_s\m{A}_s^\dagger) \m{x} \in {\rm im}(\m{I}_d-\m{A}_s\m{A}_s^\dagger) = {\rm ker}(\m{A}_s\m{A}_s^\dagger) = {\rm ker}(\m{A}_1+\m{A}_2)  \subseteq {\rm ker}(\m{A}_2)$, and it follows that $\m{A}_2[\m{I}_d-(\m{A}_1+\m{A}_2)(\m{A}_1+\m{A}_2)^\dagger]\m{x} = \m{0}_d$.
\end{itemize}
Thus, $(\m{A}_1\wedge\m{A}_2 - \m{A}_2\wedge\m{A}_1)\m{x} = \m{0}_d,~ \forall \m{x}\in\mb{R}^d$. This implies that $\m{A}_1\wedge\m{A}_2 = \m{A}_2\wedge\m{A}_1 = (\m{A}_1\wedge\m{A}_2)^\top$.

ii. By further expanding the summands in \eqref{eq:A1}, we obtain 
$\m{A}_1\wedge\m{A}_2 = \m{A}_1+\m{A}_2-\m{A}_2\m{A}_s^\dagger\m{A}_1 -\m{A}_1\m{A}_s^\dagger\m{A}_1 -\m{A}_2\m{A}_s^\dagger\m{A}_2,$
Thus, $\m{B} = \m{A}_1\wedge\m{A}_2 + (\m{A}_1\wedge\m{A}_2)^\top = \m{A}_1\wedge\m{A}_2 + \m{A}_2\wedge\m{A}_1 = \m{A}_s -\m{A}_1\m{A}_s^\dagger\m{A}_1-\m{A}_2\m{A}_s^\dagger\m{A}_2 =\m{A}_1^\top[\m{A}_1^\dagger - \m{A}_s^\dagger] \m{A}_1 + \m{A}_2^\top[\m{A}_2^\dagger - \m{A}_s^\dagger] \m{A}_2.$ 
When restricted to ${\rm im}(\m{A}_1)$, $\m{A}_1^\dagger-\m{A}_s^\dagger \succeq 0$ \citep{Horn2012matrix}. Likewise, when restricted to ${\rm im}(\m{A}_2)$, $\m{A}_2^\dagger-\m{A}_s^\dagger \succeq 0$. It follows that $\m{B} \succeq 0$ and thus, $\m{A}_1\wedge\m{A}_2 \succeq 0$.

iii. Consider equation 
\begin{align}
\m{A}_1(\m{A}_1+\m{A}_2)^{\dagger}\m{A}_2\m{x}= \m{A}_2(\m{A}_1+\m{A}_2)^{\dagger}\m{A}_1\m{x}= \m{0}_d. \label{eq:A3}
\end{align}
It is easy to see that $\m{x}\in {\rm ker}(\m{A}_2)$ and $\m{x}\in {\rm ker}(\m{A}_1)$ are solutions of \eqref{eq:A3}. Thus, ${\rm ker}(\m{A}_1) + {\rm ker}(\m{A}_2) \subseteq {\rm ker}(\m{A}_1 \wedge \m{A}_2)$. Without loss of generality, assume that $({\rm ker}(\m{A}_1) + {\rm ker}(\m{A}_2))^\perp = {\rm im}(\m{A}_1) \cap {\rm im}(\m{A}_2) \neq \{\m{0}_d\}$, and let $\m{y} \in {\rm im}(\m{A}_1) \cap {\rm im}(\m{A}_2)$. From ii., $\m{y}^\top\m{B}\m{y} = 0$ implies that $\m{A}_1=\m{A}_2 = \bm{\Theta}_d$, which is a contradiction. Thus, $({\rm im}(\m{A}_1) \cap {\rm im}(\m{A}_2)) \cap {\rm ker}(\m{A}_1 \wedge \m{A}_2) = \{\m{0}_d\}$. Hence, ${\rm ker}(\m{A}_1) + {\rm ker}(\m{A}_2) = {\rm ker}(\m{A}_1 \wedge \m{A}_2)$.

iv. This property follows directly by applying iii.

v. Since ${\rm ker}(\m{A}_1\vee\m{A}_2) \subseteq {\rm ker}(\m{A}_1)$, we have ${\rm ker}((\m{A}_1\vee\m{A}_2)\wedge\m{A}_3) \subseteq {\rm ker}(\m{A}_1\wedge\m{A}_3)$. Similarly, we have 
${\rm ker}((\m{A}_1\vee\m{A}_2)\wedge\m{A}_3) \subseteq {\rm ker}(\m{A}_2\wedge\m{A}_3)$. Thus, ${\rm ker}((\m{A}_1\vee\m{A}_2)\wedge\m{A}_3) \subseteq {\rm ker}(\m{A}_1\wedge\m{A}_3) \cap {\rm ker}(\m{A}_2\wedge\m{A}_3)={\rm ker}((\m{A}_1 \wedge \m{A}_3) \vee (\m{A}_2\wedge\m{A}_3))$.

\section{Proof of Lemma~\ref{lem:decision_operator}}
\label{app:decision}
i. This property follows directly from the definition of the decision operator. If $\m{A}_1$ is positive definite, ${\rm ker}(\m{A}_1)={\rm ker}(\m{I}_d)={\ker}(\mc{D}(\m{A}_1))$. Otherwise, the decision operator does not changes $\m{A}_1$, thus, leaves its kernel unaltered.
    
ii. Consider two possibilities as follows:
\begin{itemize}
   \item If $\exists \m{A}_i,i\in \{1,2\}$ positive definite. Then, $\m{A}_1+\m{A}_2$ is positive definite and thus, the equality holds.
   \item $\m{A}_i,i\in \{1,2\}$ are not positive definite. Then, $\mc{D}(\m{A}_1)\vee \mc{D}(\m{A}_2)=\m{A}_1 \vee \m{A}_2$ and thus, the equality holds.
\end{itemize}
    
iii. Consider two possibilities as follows:
\begin{itemize}
   \item If $\exists \m{A}_i,i\in \{1,2\}$ positive definite, without loss of generality, says $\m{A}_1$. Then, ${\rm ker}(\mc{D}(\m{A}_1)\wedge\mc{D}(\m{A}_2))={\rm ker}(\m{A}_2)={\rm ker}(\mc{D}(\m{A}_1\wedge\m{A}_2))$. The equality holds.
   \item $\m{A}_1,\m{A}_2$ are not positive definite. Then, $\mc{D}(\m{A}_1)\wedge \mc{D}(\m{A}_2)=\m{A}_1 \wedge \m{A}_2$ and thus, the equality follows from (i).
\end{itemize}
    
iv. These identities are special cases of (ii) and (iii).

\section{Proof of Lemma \ref{lem:kernelSpaces2matrix}}
\label{app:kernel2matrix}
Based on Equation~\eqref{eq:multi_vertex_condition} and the fact that the decision operator does not alter the kernel of its operand, we have
\begin{align}
{\rm ker}(\mc{R}_{ij}^1) &= {\rm ker}(\m{A}_{ij}) = {\rm ker}(\mc{D}(\m{A}_{ij})) ={\rm ker}\left([\m{A}^1]_{ij} \right),\nonumber\\
{\rm ker}(\mc{R}_{ij}^2) &= \bigcap_{k \in \mc{N}_i} ({\rm ker}(\m{A}_{ik})+{\rm ker}(\m{A}_{kj})) \nonumber \\
& = {\rm ker}\left( \bigvee_{k=1}^n (\m{A}_{ik} \wedge \m{A}_{kj}) \right) \nonumber \\
&= {\rm ker}\left( \mc{D}\left[\bigvee_{k=1}^n (\mc{D}[\m{A}_{ik}] \wedge \m{A}_{kj}) \right] \right) \nonumber\\
&= {\rm ker}\left([\m{A}^2]_{ij}\right), \label{eq:Rij2}\\
{\rm ker}(\mc{R}_{ij}^3) &= \bigcap_{k=1}^n \left(\bigcap_{k_1 =1}^n ({\rm ker}(\m{A}_{ik_1})+{\rm ker}(\m{A}_{k_1k})) + {\rm ker}(\m{A}_{kj})\right) \nonumber\\
&= {\rm ker}\left(  \bigvee_{k=1}^n \left(\bigvee_{k_1=1}^n (\m{A}_{ik_1} \wedge \m{A}_{k_1k}) \right) \wedge \m{A}_{kj}  \right) \nonumber \\ 
&= {\rm ker}\left(  \mc{D}\left(\bigvee_{k=1}^n \mc{D} \left(\bigvee_{k_1=1}^n (\mc{D}(\m{A}_{ik_1}) \wedge \m{A}_{k_1k}) \right) \wedge \m{A}_{kj} \right) \right) \nonumber \\
&= {\rm ker}\left(\mc{D}\left(\bigvee_{k=1}^n ([\m{A}^2]_{ik}\wedge \m{A}_{kj}) \right)\right)\nonumber\\ &= {\rm ker}\left([\m{A}^3]_{ij}\right). \label{eq:Rij3}
\end{align}

Keeping in mind that Equation~\eqref{eq:Rij2} is indeed Equation~\eqref{eq:multi_vertex_condition} restricted to all paths of length 2 joining $v_i,v_j$, and each $v_k,~k \in \mc{N}_i \cap \mc{N}_j$ acts as an intermediate vertex. Likewise, Equation~\eqref{eq:Rij3} applies \eqref{eq:multi_vertex_condition} twice, and $v_{k_1}$ and $v_k$ are two intermediate vertices along each path of length 3 from $v_i$ to $v_j$ (allowing loops). With this observation, for each $t\ge 2$, we have
\begin{align}
{\rm ker} (\mc{R}_{ij}^t) &= \bigcap_{k_{t-1} = 1}^n \left( \bigcap_{k_{t-2} =1}^n \left(  \ldots \bigcap_{k_1=1}^n \left( {\rm ker}(\m{A}_{ik_1}) + {\rm ker}(\m{A}_{k_1k_2})\right) + \ldots + {\rm ker}(\m{A}_{k_{t-2}k_{t-1}}) \right) + {\rm ker}(\m{A}_{k_{t-1}j}) \right) \nonumber\\
&= {\rm ker}\left(  \bigvee_{k_{t-1}=1}^n \left( \bigvee_{k_{t-2}=1}^n \left( \ldots \bigvee_{k_1=1}^n (\m{A}_{ik_1} \wedge \m{A}_{k_1k_2})  \wedge \ldots  \right) \wedge \m{A}_{k_{t-2}k_{t-1}}\right)\wedge \m{A}_{k_{t-1}j} \right) \nonumber\\
&= {\rm ker}\left( \mc{D}\left( \bigvee_{k_{t-1}=1}^n \mc{D}\left( \bigvee_{k_{t-2}=1}^n \left( \ldots \mc{D}\left(\bigvee_{k_1=1}^n (\mc{D}(\m{A}_{ik_1})  \wedge \m{A}_{k_1k_2})  \right)  \wedge \ldots  \right) \wedge \m{A}_{k_{t-2}k_{t-1}}\right)\wedge \m{A}_{k_{t-1}j} \right)\right) \nonumber\\
&= {\rm ker}\left([\m{A}^t]_{ij} \right).
\end{align}

Thus, ${\rm ker}(\mc{R}_{ij}^t) = {\rm ker}\left([\m{A}^t]_{ij}\right)$ for all $t=1,\ldots,n-1$. It has been shown from \eqref{eq:kernel_Sij_inf1} that to assess the connectivity between two vertices, we only need to consider the set of paths joining them with length up to $n-1$. When restricted to this set, and constraint all paths $\mc{P}_{k_pk_{p+1}}$ to be length 1, we obtain ${\rm ker}(\mc{Q}_{ij})=\bigcap_{t=1}^{n-1}{\rm ker} \left(\mc{R}_{ij}^t \right)$. Based on Lemma~\ref{lem:multi_intermediate_vertex}, if ${\rm dim} \bigcap_{t=1}^{r} {\rm ker}\left([\m{A}^t]_{ij} \right) = 0$, for some $r=1,\ldots,n-1$, then $v_i$ and $v_j$ must belong to a same cluster.

\section{Proof of Lemma~\ref{lem:monotonicity_kernel}}
\label{app:monotone}
i. Clearly, the output of the ``$\wedge$'' (or the ``$\vee$'') operator of two symmetric positive semidefinite matrices is again a positive semidefinite matrix. The decision operator also outputs a positive semidefinite matrix given that the operand is a positive semidefinite matrix. Thus, these three operators can be applied in sequences and the power of a block matrix $\m{A} \in \mb{R}^{dn \times dn}$ is well-defined. 
We have, 
\begin{align}
{\rm ker}\left([\m{M}(G,k)]_{ij}\right)  & = {\rm ker}\left(\left[\mc{D}\left(\m{M}(G,k-1) \vee \m{A}^k\right) \right]_{ij} \right)\nonumber\\
           &= {\rm ker}\left(\left[\mc{D}\left(\mc{D}(\m{M}(G,k-2) \vee \m{A}^{k-1}) \vee \m{A}^k\right)\right]_{ij} \right) \nonumber\\
           &= {\rm ker}\left(\left[\mc{D}\left((\mc{D}(\m{M}(G,k-2)) \vee \mc{D}(\m{A}^{k-1})) \vee \m{A}^k\right) \right]_{ij} \right)\nonumber\\
           &= {\rm ker}\left(\left[\mc{D}\left(\mc{D}(\m{M}(G,k-2)) \vee \m{A}^{k-1} \vee \m{A}^k\right)\right]_{ij} \right) \nonumber\\
           &\qquad\qquad \vdots \nonumber \\
           &= {\rm ker}\left(\left[\mc{D}\left(\mc{D}(\m{M}(G,1)) \vee \m{A}^2\vee\ldots \vee \m{A}^{k-1} \vee \m{A}^k\right)\right]_{ij} \right) \nonumber\\
           &={\rm ker}\left(\left[\mc{D}\left(\bigvee_{i=0}^k\m{A}^k\right)\right]_{ij}\right). \label{eq:appD}
\end{align}

ii. We have $[\m{M}(G,k)]_{ij}=\mc{D}([\m{M}(G,k-1)]_{ij} \vee \mc{D}([\m{A}^k]_{ij}))$. 
It follows from Lemma~\ref{lem:series_sum}(iii) that ${\rm ker}([\m{M}(G,k)]_{ij}) = {\rm ker}(\mc{D}([\m{M}(G,k-1)]_{ij}) \cap {\rm ker}(\mc{D}([\m{A}^k]_{ij}))) \subseteq {\rm ker}(\mc{D}([\m{M}(G,k-1)]_{ij})$, which completes the proof. 

\end{document}